\documentclass{article}

\usepackage{PRIMEarxiv}

\usepackage[utf8]{inputenc} 
\usepackage[T1]{fontenc}    
\usepackage{hyperref}       
\usepackage{url}            
\usepackage{booktabs}       
\usepackage{amsfonts}       
\usepackage{nicefrac}       
\usepackage{microtype}      
\usepackage{lipsum}
\usepackage{fancyhdr}       
\usepackage{graphicx}       
\usepackage{todonotes}
\usepackage{amsmath}
\usepackage{float}
\usepackage{algpseudocode}
\usepackage{algorithm}
\usepackage{multirow}
\usepackage{subcaption}
\graphicspath{{media/}}     

\newcommand{\context}[1]{\textcolor{black}{#1}}
\newcommand{\challenge}[1]{\textcolor{black}{#1}}
\newcommand{\hypothesis}[1]{\textcolor{black}{#1}}
\newcommand{\proposal}[1]{\textcolor{black}{#1}}
\newcommand{\evaluation}[1]{\textcolor{black}{#1}}
\newcommand{\paperdesc}[1]{\textcolor{black}{#1}}
\newcommand{\rewrite}[1]{\textcolor{black}{#1}}
\newcommand{\paco}[1]{{#1}}

\title{Resilient Federated Chain: Transforming Blockchain Consensus into an Active Defense Layer for Federated Learning
}

\author{
  Mario García-Márquez, Nuria Rodríguez-Barroso \\
  Department of Computer Science and Artificial Intelligence,\\ Andalusian Research Institute in Data Science \\
  and Computational Intelligence (DaSCI) \\
  University of Granada \\
  Granada\\
  \texttt{\{mariogmarq, rbnuria\}@ugr.es} \\
   \And
  M.V. Luzón \\
    Department of Software Engineering,\\ 
    Andalusian Research Institute in Data Science \\
  and Computational Intelligence (DaSCI) \\
  University of Granada \\
  Granada\\
  \texttt{luzon@ugr.es} \\
  \And
  Francisco Herrera \\
  Department of Computer Science and Artificial Intelligence,\\ Andalusian Research Institute in Data Science \\
  and Computational Intelligence (DaSCI) \\
  University of Granada \\
  Granada\\
  \texttt{herrera@decsai.ugr.es} \\
}

\begin{document}
\maketitle

\begin{abstract}
Federated Learning (FL) has emerged as a key paradigm for building Trustworthy AI systems by enabling privacy-preserving, decentralized model training. \rewrite{However, FL is highly susceptible to adversarial attacks that compromise model integrity and data confidentiality, a vulnerability exacerbated by the fact that conventional data inspection methods are incompatible with its decentralized design.} While integrating FL with Blockchain technology has been proposed to address some limitations, its potential for mitigating adversarial attacks remains largely unexplored. This paper introduces \rewrite{Resilient Federated Chain (RFC)}, a novel \rewrite{blockchain-enabled} FL framework designed specifically to enhance resilience against such threats. RFC \rewrite{builds upon the} existing Proof of Federated Learning architecture by repurposing the redundancy \rewrite{of} its Pooled Mining mechanism as an \rewrite{active} defense layer that can be combined with robust aggregation rules. Furthermore, the framework introduces a flexible evaluation function in its consensus mechanism, allowing for adaptive defense against different attack strategies. \rewrite{Extensive experimental evaluation on image classification tasks under various adversarial scenarios, demonstrates that RFC significantly improves robustness compared to baseline methods, providing a viable solution for securing decentralized learning environments.}
\end{abstract}

\section{Introduction}\label{sec:intro}
\context{The rapid integration of Artificial Intelligence (AI) systems into critical sectors \cite{islam2022systematic}, including healthcare, finance, and law enforcement, has heightened the demand for comprehensive frameworks that ensure conformity with ethical, legal, and societal expectations. In this context, the paradigm of trustworthy AI (TAI) \cite{li2023trustworthy} has emerged as a central principle guiding both the development and the responsible deployment of AI technologies. The European Commission, among other influential institutions, has codified this paradigm in its Ethical Guidelines, where TAI is delineated through seven fundamental requirements \cite{diaz2023connecting}: (1) human agency and oversight, (2) technical robustness and safety, (3) privacy and data governance, (4) transparency, (5) diversity and fairness, (6) societal well-being, and (7) accountability.  Within this normative and technical framework, Federated Learning (FL) \cite{kairouz2021advances} has emerged as a particularly relevant machine learning (ML) approach. FL enables decentralized model training across multiple clients while ensuring that raw data remains localized, thereby addressing pressing concerns regarding data privacy and security. Beyond its immediate technical advantages, FL constitutes a practical mechanism for embedding the principles of trustworthy AI into operational systems, effectively bridging theoretical guidelines with applied ML practice \cite{li2021survey}.}

\challenge{Nonetheless, it is essential to recognize that, despite its inherent \rewrite{privacy preserving} properties, FL remains vulnerable to adversarial threats capable of undermining both data confidentiality and model integrity. Adversarial attacks in ML are primarily designed either to degrade model performance or to compromise sensitive information~\cite{rodriguez2023survey}. Given that FL represents a particular instantiation of distributed ML, it is subject to these same risks. A significant subset of such attacks arises from the deliberate manipulation of training data, \rewrite{leading to corrupted updates}. In the context of FL, where raw data \rewrite{remains} inaccessible to central entities, conventional inspection-based techniques for detecting data tampering cannot be employed. Consequently, a notable limitation of FL lies in its susceptibility to adversarial interventions that compromise the reliability of the learning process, \rewrite{requiring specialized mitigation at the update level. The complexity of these threats is further amplified in decentralized federated learning, where the lack of a central authority necessitates more sophisticated defense strategies \cite{hallaji2024decentralized}.}}

\hypothesis{To mitigate these risks without compromising privacy, the integration of FL with ledger technologies such as Blockchain has intensified~\cite{cai2025blockchain}. While promising, the majority of existing academic literature emphasizes dimensions such as scalability, single points of failure, and incentive mechanisms. Crucially, few works have systematically leveraged the pooled consensus architecture itself as an \rewrite{active} defense layer against the data tampering described above.}

\proposal{To bridge this gap, we propose Resilient Federated Chain (RFC), a framework designed to employ mining redundancy as a robust defensive mechanism. Grounded in the Proof of Federated Learning (PoFL)~\cite{qu2021proof} architecture, which utilizes disjoint mining pools primarily for consensus generation, \rewrite{we hypothesize this redundancy of mining pools as a method for adversarial robustness}. Our approach \rewrite{takes advantage of this property in order to tailor the architecture as an active defense layer for adversarial attacks}. By treating the parallel model generation across pools as a form of competitive validation, RFC can isolate and discard corrupted updates that emerge from malicious clients. Furthermore, we extend the architecture to address the vulnerability of standard aggregation rules to outliers. We introduce a modular consensus design where the cost function and aggregation mechanism are rendered interchangeable. This flexibility provides an \rewrite{active} defense layer capable of mitigating attacks that would otherwise bypass rigid, accuracy-based selection criteria, extending the resilience of PoFL.}

\proposal{\rewrite{Critically, RFC differs fundamentally from traditional ensemble or committee-based FL approaches~\cite{che20222decentralized}}. In standard \rewrite{committee-based} FL, a central aggregator or a fixed voting committee still represents a single point of failure and trust. If the aggregator is compromised, the defense collapses. RFC removes this reliance by repurposing the mining redundancy of PoFL. We demonstrate that the 'wasteful' parallel computation inherent to Proof of Work can be transformed into a probabilistic guarantee of Byzantine Fault Tolerance. By decoupling the defense from any central authority and embedding it into the cryptographic consensus itself, RFC ensures that the \rewrite{active defense layer} is a mathematical function of pool diversity rather than an assumption of server integrity.}

\evaluation{To evaluate RFC, we design a systematic experimental framework that examines both its performance and robustness across diverse conditions. The evaluation involves training image classification models on multiple FL datasets of varying complexity, ensuring that the approach is tested under realistic and heterogeneous data distributions. We further assess resilience against adversarial threats---the property of the federated environment of not being affected by those threats---by simulating both Byzantine and backdoor attacks~\cite{surveynuria}, which represent common challenges in FL environments. Comparative analyses are conducted against established baseline aggregation rules to determine the relative benefits of RFC. Finally, experiments are carried out under different scenarios, including benign settings and adversarial conditions, to provide a comprehensive understanding of the framework’s behavior in practical deployments.}

\paperdesc{The rest of this paper is structured as follows. Section~\ref{sec:background} provides the necessary background and related work, starting with an introduction to FL in Subsection~\ref{sec:fl}, an introduction to Blockchain in Subsection~\ref{sec:blockchain}, followed by the integration of Blockchain in this paradigm in Subsection~\ref{sec:blockfed}, and an overview of the PoFL consensus mechanism in Subsection~\ref{sec:pofl}. In Section~\ref{sec:proposal}, we introduce our proposed framework, RFC, which is designed to improve the resilience of FL systems. The experimental setup is detailed in Section~\ref{sec:setup}, where we describe the datasets and models in Subsection~\ref{sec:datasets}, the adversarial attacks implemented in Subsection~\ref{sec:attacks}, the baselines and scenarios for evaluation in Subsection~\ref{sec:baselines}, the evaluation metrics being considered in Subsection~\ref{sec:metrics}, and the implementation details in Subsection~\ref{sec:details}. Section~\ref{sec:analysis} presents and analyzes the results of our experiments. Finally, Section~\ref{sec:conclusions} concludes the paper and outlines future work.}

\section{Background and Related Work}\label{sec:background}
\rewrite{This section \rewrite{provides sufficient} context to properly understand and frame the contributions of this paper. First, an introduction to FL is given in Section~\ref{sec:fl}, \rewrite{followed} by an explanation of the core concepts of Blockchain in Section~\ref{sec:blockchain}. Afterwards, building on these concepts, we provide a view of the current state of integrating Blockchain technologies with FL and relevant related work to this manuscript in Section~\ref{sec:blockfed}. Finally, we provide a deeper dive into a core proposal for this integration, which is Proof of Federated Learning, explained in Section~\ref{sec:pofl}.}

\subsection{Introduction to Federated Learning. Why and How?}\label{sec:fl}
The increasing quantity and variety of data have created significant problems for data privacy and the effective processing of massive datasets. FL has arisen as a promising methodology to address these challenges, focusing on privacy, \rewrite{communication overhead}, and data accessibility.

\rewrite{Traditionally, centralized machine learning necessitates the aggregation of user data on a single server, which inherently increases the attack surface for privacy violations \cite{Shokri2016MembershipIA}. This risk is particularly \rewrite{evident} in sensitive domains such as healthcare and finance \cite{abouelmehdi2017big}, and is further complicated by legal frameworks like the General Data Protection Regulation (GDPR) \cite{gdpr2}, which \rewrite{demands} privacy-preserving AI architectures. Beyond privacy, the sheer scale of modern datasets, exacerbated by the proliferation of \rewrite{the Internet of Things}, makes the transfer of raw data to a central hub computationally expensive and sensitive to latency \cite{bib:mcmahan16communicationefficient}. Furthermore, data is often geographically and institutionally fragmented. These "data silos" are reinforced by legal and technical barriers that prevent unified access, rendering traditional centralized training nearly impossible \cite{abouelmehdi2017big}.}

\rewrite{To navigate these constraints, FL \cite{yang2019federated} introduces a distributed learning framework that enables the collaborative optimization of a global model without requiring participants to disclose their local data. The process is orchestrated across a network of clients, typically operating through two primary phases:}

\begin{enumerate}
    \item \textit{Model Training Phase}: Each client trains a local model on its own private data and shares only the resulting model updates, not the data itself. These local models are then aggregated to form a global model, ensuring data privacy is maintained throughout the learning procedure.
    \item \textit{Inference Phase}: The combined global model is then utilized to perform predictions on new data.
\end{enumerate}

\rewrite{These phases can be executed synchronously or asynchronously, depending on the network topology and data availability. Beyond the technical objective of privacy, the sustainability of such a system often relies on fair value distribution mechanisms to ensure that the benefits of the collaborative model are shared equitably among all participating clients.}

Formally, an FL setup can be described as follows: Let there be a collection of clients or data holders $\{C_1, \dots, C_n\}$, where each holds local training data $\{D_1, \dots, D_n\}$. Each client $C_i$ possesses a local model defined by parameters $L_i$. The main goal of FL is to train a global model, defined by parameters $G$, by leveraging the distributed data from all clients through an iterative method known as a learning round. In each learning round $t$:

\begin{enumerate}
    \item Each client trains its local model on its own local data $D^t_i$, thereby updating its parameters from $L^{t}_i$ to $\hat{L}^t_i$.
    \item The global parameters $G^t$ are determined by aggregating the updated local models $\{\hat{L}^t_1, \dots, \hat{L}^t_n\}$ using a specific federated aggregation operator $\Delta$:
    \begin{equation}
    \begin{split}
        G^t = \Delta(\hat{L}^t_1,\hat{L}^t_2, \dots, \hat{L}^t_n)\\
        L^{t+1}_i \leftarrow G^t, \quad \forall i \in \{1, \dots, n\}.
    \end{split}
    \label{eq_fl_aggregation}
    \end{equation}
\end{enumerate}

\begin{figure}[!ht]
    \centering
    \includegraphics[width=0.9\linewidth]{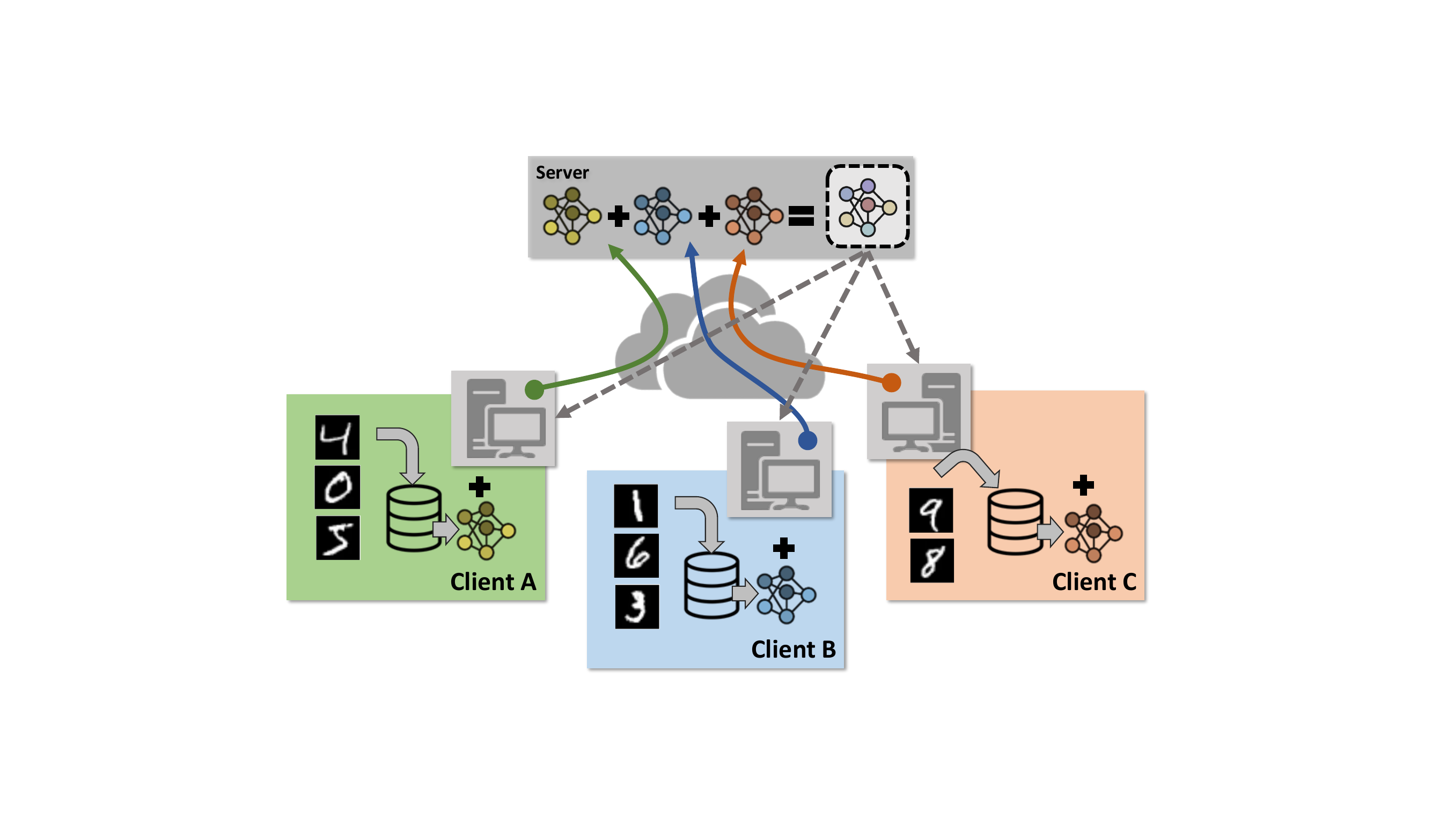}
    \caption{Descriptive diagram showing the architecture of a FL system. Source: \cite{luzon2024tutorial}.}\label{fig:fl}
\end{figure}

This repetitive refinement process continues until a specified stopping condition is met, leading to a global model parametrized by $G$ that captures the collective insights of all participating entities. This is visually depicted in Figure~\ref{fig:fl}.

However, since FL is a specialized instance of ML, inherits the susceptibility of its parent field to adversarial attacks that seek to degrade performance or compromise privacy. The existing literature categorizes these attacks through several criteria~\cite{surveynuria}, including the attacker's knowledge of the system, the manipulation of model behavior, and the attack's objective. Within the latter taxonomy, untargeted attacks aim solely to reduce the model's performance on the primary learning task. A particularly challenging case of this attack is the Byzantine attack~\cite{byzantinegenerals, Hu2021ChallengesAA}, in which a subset of clients submit arbitrary updates. These updates are typically generated randomly or derived from models trained on manipulated data, effectively producing random updates.

These attacks are frequently used in conjunction with model replacement techniques~\cite{howtobackdoor}. This is due to the fluctuating proportion of adversarial clients, which can prevent the mitigation of malicious updates, as the sheer number of benign client updates may not effectively counteract the influence of compromised updates.

\subsection{Core Blockchain Concepts}\label{sec:blockchain}

At its core, a Blockchain is a decentralized, distributed, and immutable digital ledger used to record transactions or any digital interaction across many computers~\cite{tran2022handbook}. The technology is designed so that anything recorded cannot be altered retroactively without the alteration of all subsequent blocks and the consensus of the network. This resistance to modification is one of its key security features.

The ledger is structured as a chronologically ordered chain of records called blocks. Each block, denoted as $B_i$ for the $i$-th block in the chain, contains several key components~\cite{nakamoto2008bitcoin}:
\begin{itemize}
    \item A batch of data, $D_i$.
    \item A timestamp, $T_i$, marking its creation time.
    \item A cryptographic hash of the current block's contents, $H_i$.
    \item The cryptographic hash of the preceding block, $H_{i-1}$.
\end{itemize}

This chain is formed by cryptographically linking each block to the one before it. A cryptographic hash function, $H(\cdot)$, is a mathematical algorithm that maps input data of any size to a fixed size string of bytes. A crucial property of such functions is that they are deterministic and computationally infeasible to invert. The hash of a block $B_i$ is generated from its contents, including the hash of the previous block, $H_{i-1}$. This can be formally expressed as:

\begin{equation}
    H_i = H(D_i, T_i, N_i, H_{i-1})
\end{equation}

where $N_i$ is a special number known as a nonce. This recursive inclusion of the previous block's hash creates an unbreakable and tamper evident chain. If an attacker were to alter the data $D_k$ in a past block $B_k$, its hash $H_k$ would change. Consequently, the hash $H_{k+1}$ of the next block $B_{k+1}$ would be incorrect (as it was computed using the original $H_k$), invalidating it and all subsequent blocks in the chain.

For a new block to be added to the chain, the nodes in the decentralized network must agree on its validity. This agreement is achieved through a so called consensus mechanism. The original and most well-known consensus mechanism is Proof of Work (PoW)~\cite{nakamoto2008bitcoin}. In a PoW system, network participants, known as miners, compete to solve a computationally intensive puzzle. The first miner to solve the puzzle gets to add the next block to the blockchain and is typically rewarded. The puzzle involves finding a specific value for the nonce, $N_i$, such that when it is included with the other block data, the resulting block hash $H_i$ meets a certain condition. This condition is usually that the hash value must be below a specific target value, $T_{target}$, which is adjusted periodically to control the rate of block creation. The mining process is a brute-force search for a valid nonce as in Equation~\ref{eq:nonce}.

\begin{equation}\label{eq:nonce}
    \text{Find } N_i \text{ such that } H(D_i, T_i, N_i, H_{i-1}) < T_{target}
\end{equation}

The only way to find a suitable $N_i$ is to try different values iteratively until a valid hash is produced. This process requires an immense amount of computational power and, by extension, electrical energy. While this computational expense is what secures the network, making it prohibitively costly for an attacker to overpower, it is also its primary drawback. The vast computational resources expended in PoW mining do not produce any intrinsically useful output beyond securing the ledger. This perceived wastefulness has driven the development of alternative consensus mechanisms.

\subsection{Blockchain Integration in Federated Learning and Related Work}\label{sec:blockfed}
The integration of Blockchain technology and FL is becoming a prominent research field, addressing inherent challenges within traditional FL systems. This integration is motivated by several critical factors, primarily focusing on enhancing security, privacy, decentralization, and accountability in collaborative AI.

A primary driver for integrating Blockchain with FL is the substantial enhancement of security and privacy. Traditional FL architectures often deal with vulnerabilities such as data tampering, unauthorized access, and the susceptibility to single points of failure \cite{kairouz2021advances}. Blockchain's decentralized ledger inherently mitigates these risks. The immutability and traceability features of a Blockchain ensure the integrity and authenticity of model updates, providing a robust defense against malicious alterations and ensuring that all participants can verify the provenance of shared models \cite{qammar2023securing}.

Furthermore, the integration fundamentally transforms the architectural paradigm of FL by eliminating the reliance on a central server, which typically represents a single point of failure and a bottleneck for scalability \cite{kairouz2021advances}. Blockchain facilitates a truly decentralized FL framework, significantly improving system reliability and robustness. This distributed approach inherently addresses issues such as unreliable model uploads and heterogeneity across participating nodes, fostering a more resilient distributed learning environment \cite{li2022blockchain}.

Blockchain technology is also instrumental in establishing transparent and equitable incentive mechanisms within FL ecosystems. These mechanisms are crucial for motivating participants to contribute high-quality data and model updates, which is essential for the effectiveness and fairness of the collaborative learning process. The inherent transparency of Blockchain allows for the verifiable and automated distribution of rewards, ensuring that contributions are appropriately recognized and compensated \cite{qammar2023securing}.

Blockchain technology \rewrite{has also} been employed in order to mitigate adversarial attacks. For example, \cite{attiaou2024blockchain} \rewrite{proposed} a Blockchain architecture which makes use of a Proof of Stake consensus mechanism in order to avoid \rewrite{a} single point of failure and protect against poisoning attacks. \cite{dong2024defending} also \rewrite{defends} against poisoning attacks by incorporating \rewrite{peer-to-peer} voting and a \rewrite{reward-and-slash} mechanism into the architecture. \rewrite{Furthermore, recent work has explored the use of blockchain for auditing and verifying the behavior of aggregators in decentralized environments \cite{hallaji2025trustchain}.}

\subsection{Proof of Federated Learning}\label{sec:pofl}
\rewrite{To effectively integrate Blockchain into the FL paradigm, the choice of consensus mechanism is paramount. Historically, PoW~\cite{nakamoto2008bitcoin} established the foundation for decentralized trust. However, its reliance on solving cryptographically hard but otherwise purposeless hash puzzles results in massive energy consumption and computational waste. To address this inefficiency, the concept of Proof of Useful Work (PoUW)~\cite{ball2018proofs} was introduced. PoUW proposes replacing the arbitrary hashing tasks of PoW with computationally intensive problems whose solutions provide external value, such as scientific simulations or combinatorial optimization, thereby repurposing the network's energy for productive ends.}

\rewrite{Building upon this trajectory, Proof of Federated Learning (PoFL)~\cite{qu2021proof} emerges as a specialized improvement tailored for collaborative AI. PoFL refines the PoUW philosophy by specifically leveraging the stochastic and resource-heavy nature of machine learning model training as the underlying work for consensus. By employing a Pooled Mining approach, PoFL allows the network to achieve agreement through the actual iterative updates required by the FL process. This effectively converges the act of securing the ledger with the act of training the global model, creating a synergistic architecture that is both energy efficient and intrinsically aligned with the objectives of decentralized learning.}

\begin{figure}[H]
    \centering
    \includegraphics[width=0.9\linewidth]{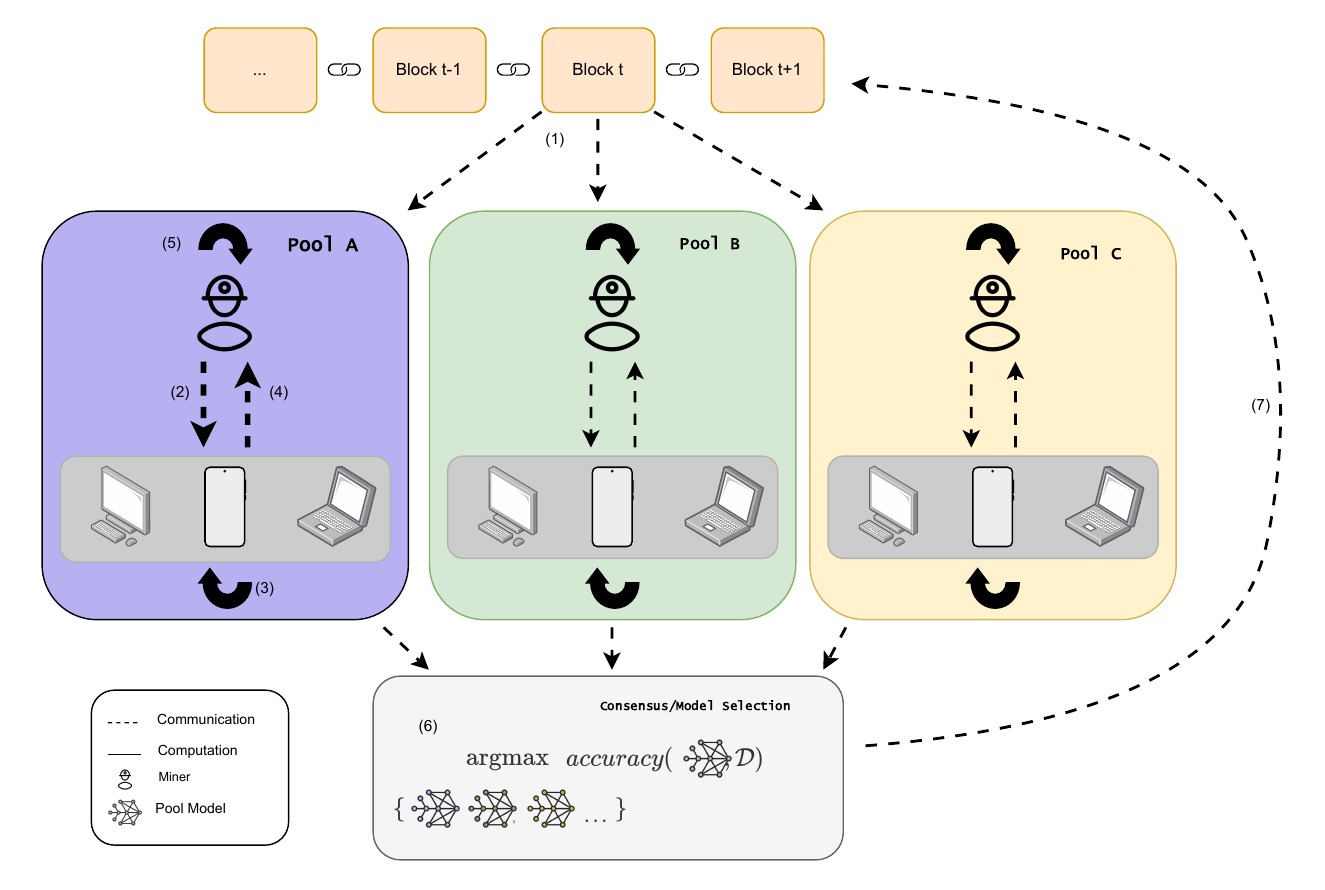}
    \caption{Schematic of the pooled mining architecture. By clustering clients into independent pools, this framework enhances system scalability and reduces blockchain communication overhead. Furthermore, the accuracy-based selection mechanism incentivizes the contribution of high quality model updates, ensuring the robustness of the global model against low-quality or malicious local contributions.}\label{fig:pooled}
\end{figure}

\rewrite{In the proposed pooled mining architecture, the network is partitioned into distinct clusters termed pools. Each pool is coordinated by a central miner and operates in isolation, ensuring no cross-pool communication occurs during the local training phase. The system follows an iterative process where pools compete to provide the most accurate global update. This workflow, illustrated in Figure \ref{fig:pooled}, unfolds in the following sequence during each round:}

\begin{enumerate}
    \item \rewrite{Global Model Retrieval: Each miner fetches the current global model from the most recent block on the blockchain.}
    \item \rewrite{Local Distribution: The miner broadcasts this global model to all participating clients within its specific pool.}
    \item \rewrite{On-device Training: Clients perform local training on their private datasets.}
    \item \rewrite{Update Collection: Clients transmit their updated local parameters back to their respective pool miner.}
    \item \rewrite{Intra pool Aggregation: The miner aggregates these updates using the FedAvg algorithm to generate a new candidate model.}
    \item \rewrite{Consensus and Evaluation: A blockchain consensus mechanism evaluates the candidate models from all competing miners. The model achieving the highest accuracy on a public validation set is selected.}
    \item \rewrite{Block Commitment: The winning miner appends the selected model to the blockchain, establishing the new global state and triggering the start of the next round.}
\end{enumerate}

\section{Resilient Federated Chain (RFC): Framework to Improve FL's Resilience}\label{sec:proposal}
The previous sections have established that while FL is a key paradigm for \rewrite{privacy-preserving} AI, it remains highly susceptible to adversarial attacks that undermine model integrity. Although the integration of Blockchain, particularly through the PoFL consensus mechanism, has been proposed to address some \rewrite{of} FL's limitations, its architecture presents an untapped potential for enhancing security.

We hypothesize that pooled mining can significantly improve the robustness of FL systems against adversarial attacks. The architecture of PoFL, originally designed as a consensus mechanism, inherently introduces network redundancy. This structural property is valuable for defense: since there is no \rewrite{inter-pool} communication during training, the negative influence of an adversarial client is confined to its specific pool, leaving the remaining pools unaffected.

Motivated by this insight, we introduce the \textbf{Resilient Federated Chain (RFC)} framework, a robust \rewrite{blockchain-enabled} meta-architecture for FL. At each training round, RFC selects the single \rewrite{best-performing} model based on a chosen evaluation metric, without imposing traditional constraints like differentiability. This approach has profound implications: it allows for the use of any arbitrary, \rewrite{non-differentiable} quality metric for model selection. Consequently, the framework becomes highly adaptable to diverse and complex problems, such as enhancing fairness, privacy, or robustness, where the objective function may not be easily expressed in a differentiable form.

\begin{figure}[!hb]
    \centering
    \includegraphics[width=0.9\linewidth]{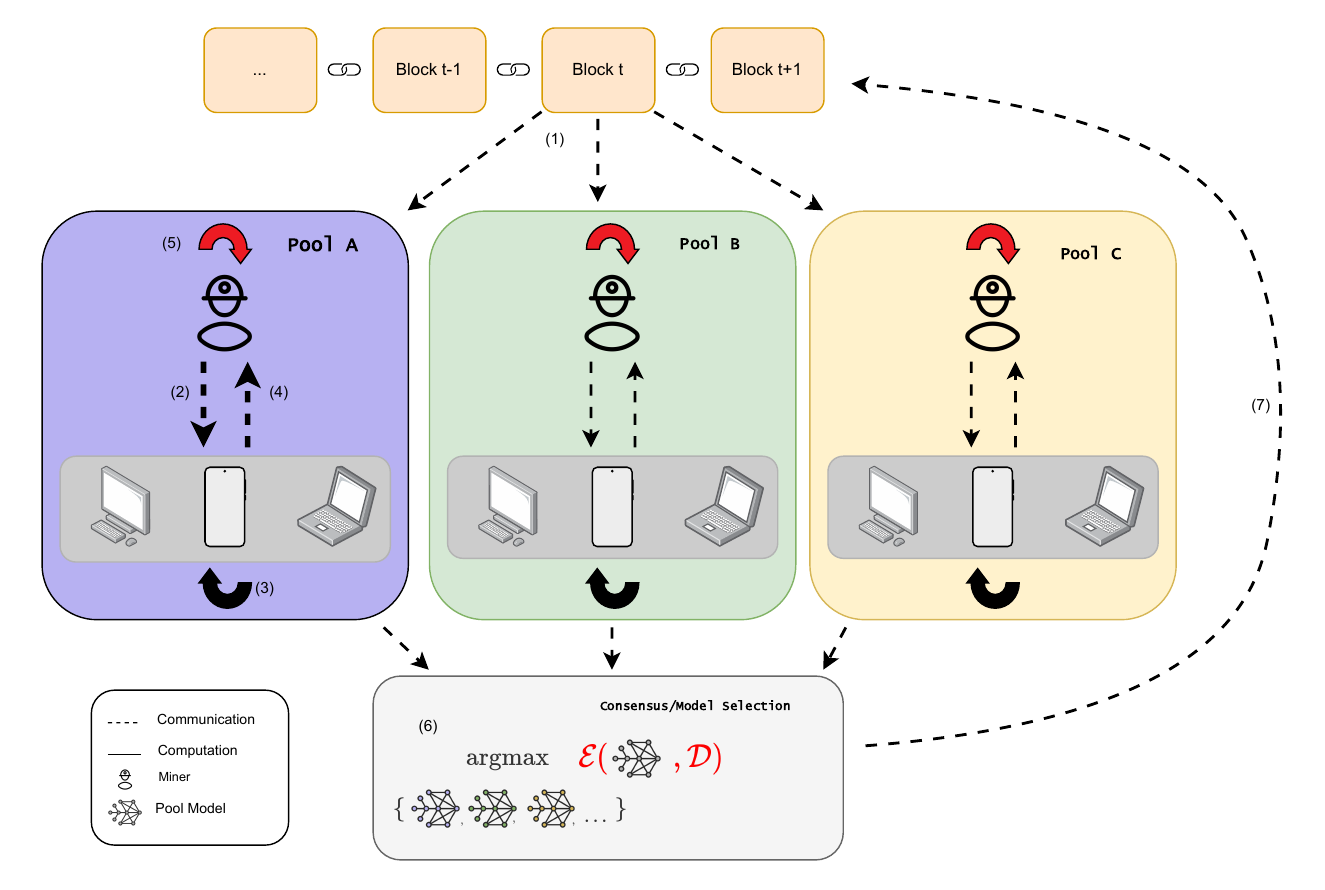}
    \caption{Diagram showing the architecture of the proposed RFC framework. \rewrite{Components} diverging from the PoFL structure are marked in red. More precisely, the aggregation algorithm employed and the metric used to select the best model in the consensus are treated as hyperparameters, allowing for a more flexible and robust architecture.}
    \label{fig:proposal}
\end{figure}

Our framework extends the foundational \rewrite{resilience} of PoFL through two main contributions. 
\begin{itemize}
\item The first is the generalization of the model evaluation criterion. While PoFL relies solely on accuracy, this choice can be suboptimal. Accuracy may conceal issues like overfitting, where loss increases while accuracy \rewrite{improves}, or produce misleading results on imbalanced validation datasets.

\rewrite{In our proposal}, the choice of the evaluation metric $\mathcal{E}$ is therefore critical and \rewrite{problem-dependent}. For instance, in scenarios with class imbalance, metrics like \rewrite{the} F1-score or AU-ROC should be preferred to prevent the consensus mechanism from selecting a model that performs poorly on minority classes. By treating the metric as a tunable hyperparameter, RFC allows practitioners to align the selection process directly with the specific goals of the learning task.

\begin{algorithm}[!ht]
\caption{Resilient Federated Chain (RFC) Training Procedure}
\label{alg:xfederated}
\begin{algorithmic}[1]
\Require Number of pools $M$, local datasets $\{D_i\}$, evaluation metric $\mathcal{E}$, aggregation rule $\mathcal{R}$
\State Initialize models $L^0$ in the first Blockchain block
\For{each round $t = 1, 2, \dots, T$}
    \For{each pool $i = 1, \dots, M$ in parallel}
        \For{each client $j$ in pool $i$}
            \State Train local model on $D_j$ and obtain update $\hat{L}_{i,j}^{t}$
        \EndFor
        \State Aggregate updates in pool $i$: 
        \[
            L^t_i \gets \mathcal{R}(\{ \hat{L}_{i,j}^{t} \}_j)
        \]
    \EndFor
    \For{each pool $i = 1, \dots, M$}
        \State Evaluate $L_i^t$ using metric $\mathcal{E}$ on validation data
    \EndFor
    \State Select best-performing pool model:
    \[
        L^t \gets \arg\max_{i} \, \mathcal{E}(L_i^t)
    \]
    \State Broadcast $L^t$ to all pools as reference for next round by appending it to the Blockchain
\EndFor
\State \Return Final global model $L^t$
\end{algorithmic}
\end{algorithm}

\item Second, we incorporate robust aggregation operators within each pool to counter malicious updates directly at the source. Rather than relying solely on the final model selection to filter out poor performers, these operators provide an initial layer of defense during gradient aggregation.

By incorporating operators like Krum, which \rewrite{offers} strong theoretical \rewrite{guarantees} of \rewrite{its} robustness against malicious participants, RFC adds a layer of provable resilience within each pool. This \rewrite{intra-pool} defense is critical in scenarios where adversarial clients are present in \emph{every} pool. In such cases, the \rewrite{inter-pool} redundancy offered by the PoFL architecture alone is insufficient. RFC’s robust aggregators ensure a baseline of resilience even if the attack is pervasive, mitigating the risk of malicious updates corrupting all candidate models before the final consensus step.
\end{itemize}

\rewrite{Figure~\ref{fig:proposal} depicts the high-level architecture of the RFC framework. While retaining the structural redundancy of PoFL via disjoint mining pools to ensure fault isolation, our design introduces critical modularity. Unlike the rigid structure of the baseline, RFC decouples the consensus logic from fixed implementation details. Specifically, the intra-pool aggregation mechanism and the consensus selection metric are treated as tunable hyperparameters rather than static rules. This architectural divergence transforms the passive redundancy of traditional ledger-based FL into a dynamic defense system, allowing the framework to adapt its robust configuration according to the specific threat model.}

\rewrite{Algorithm~\ref{alg:xfederated} formalizes this procedure, where training is distributed across M parallel pools to ensure redundancy. Crucially, the framework abstracts the defensive logic into two customizable functions: the aggregation rule $\mathcal{R}$, which filters intra-pool anomalies locally, and the evaluation metric $\mathcal{E}$, which governs the global consensus selection. By committing only the model that \rewrite{optimizes} the metric to the ledger, the system ensures that the global state evolves strictly according to the specific robustness or performance criteria defined by the network configuration.}


\section{Experimental Setup}\label{sec:setup}
In this section, we present the experimental setup used to validate the proposed RFC framework's viability and effectiveness. To evaluate the architecture, we utilize image classification models trained on various FL datasets. This section provides a comprehensive overview of our experimental setup, ensuring full reproducibility. We describe the datasets and deep learning models in Section \ref{sec:datasets}, the backdoor and Byzantine attacks in Section \ref{sec:attacks}, the threat model considered in Section \ref{sec:threat}, the baselines in Section \ref{sec:baselines}, the evaluation metrics employed in Section \ref{sec:metrics}, and relevant implementation details in Section \ref{sec:details}.

\subsection{Datasets and models}\label{sec:datasets}
In order to rigorously assess the effectiveness of our proposed approach, we employ several well-established image classification datasets. The datasets utilized are detailed below:
\begin{itemize}
    \item \textbf{CIFAR-10}. The CIFAR-10 dataset~\cite{cifar10}, a labeled subset of the extensive 80 million tiny images collection, consists of 60,000 32$\times$32 color images distributed across 10 categories. In our experiments, the training data are uniformly allocated among 200 clients.
    \item \textbf{Fashion MNIST}. The Fashion MNIST dataset~\cite{fashionmnist-2017} is designed as a more challenging benchmark compared to the original MNIST dataset. It comprises 28x28 grayscale images of various clothing items spanning 10 distinct categories. For our study, the dataset is partitioned among 200 clients.
    \item \textbf{EMNIST}. The EMNIST dataset~\cite{emnist}, introduced in 2017, extends the original MNIST dataset~\cite{lecun-1998}. We focus on the EMNIST Digits subset, which offers a balanced collection of handwritten digit samples. The data are distributed among 200 clients under an IID partitioning scheme.
    \item \textbf{Celeba-S}. The CelebA dataset~\cite{celeba} contains facial images of celebrities annotated with 40 binary attributes. For this work, we adapt it into a binary classification task by focusing on the \textit{Smiling} attribute. \rewrite{The dataset is distributed according to the identity of the celebrity, each client holding only pictures of the same person, leading to a natural non-IID distribution.}
\end{itemize}

The selected datasets represent a diverse set of challenges pertinent to FL, varying in complexity from relatively simple grayscale images to high-resolution, real-world facial images, and encompassing both multi-class (10 classes) and binary classification tasks. This diversity, summarized in Table \ref{tab:datasets_summary}, enables a thorough evaluation of the proposed method in terms of both performance and robustness.

\begin{table}[H]
    \centering
    \begin{tabular}{lll}
        \toprule
        \textbf{Dataset} & \textbf{Data Description} & \textbf{Classes} \\
        \midrule
        CIFAR-10 & 32$\times$32 Color & 10 \\
        Fashion MNIST & 28$\times$28 Grayscale & 10 \\
        EMNIST & 28$\times$28 Grayscale & 10 \\
        Celeba-S & High-resolution Color  & 2 (Binary) \\
        \bottomrule
    \end{tabular}
    \caption{Overview of the experimental datasets and their key characteristics.}\label{tab:datasets_summary}
\end{table}

Regarding the models, we employ a convolutional neural network (CNN) as the global model for the Fashion MNIST and EMNIST datasets. For CIFAR-10 and Celeba-S, we utilize a pre-trained \texttt{EfficientNet-B0}~\cite{tan2019efficientnet} as the global model. Both architectures are trained using the Adam optimizer with a learning rate of 0.001, running 10 local epochs per client in each round for 100 rounds. All the experiments have been run with 75 clients per round.

The CNN architecture designed for EMNIST consists of a feature extraction module and a classification head. The feature extraction module incorporates two consecutive convolutional blocks. The first block applies a convolutional layer with 32 filters of size 3x3 to the single-channel input, followed by a non-linear activation function. The second block comprises a convolutional layer with 64 filters of size 3x3, also followed by an activation function and a 2x2 max-pooling layer to reduce spatial dimensions. The flattened representation from the feature extraction module is subsequently passed to the classification head, which includes a dense layer with 128 units and a non-linear activation, concluding with a dense layer that projects the learned features to the corresponding number of output classes.

\subsection{Byzantine and Backdoor Attacks}\label{sec:attacks}
To evaluate the robustness of the proposed approach, we additionally consider two widely recognized adversarial threats in FL: a Byzantine attack (labelflip) and a backdoor attack (fixed pattern). The Byzantine attack involves intentional mislabeling, which causes adversarial clients to produce updates that are effectively indistinguishable from random noise. The backdoor attack, in contrast, introduces a fixed trigger pattern into input images (in our case, a white box on the image's bottom right corner) and systematically assigns this pattern to a target label, thereby embedding a hidden malicious behavior into the model while preserving its accuracy on the primary task.

Such attacks are frequently combined with a model replacement strategy, designed to amplify the effect of adversarial updates. Considering the model parameter aggregation rule defined as:

\begin{equation}\label{eq:updaterule}
    {V_\mathcal{G}}^{t+1}={V_\mathcal{G}}^{t}+ \frac{\eta}{n}\sum_{i=1}^n({V}_i^t - {V_\mathcal{G}}^t)
\end{equation}

where $\eta$ denotes the server learning rate (set to one throughout this work), an adversarial client may transmit the following scaled update:

\begin{equation}\label{eq:boost}
    \hat{{V}}_{adv}^t = \beta({V}_{adv}^t - {V_\mathcal{G}}^t)
\end{equation}

with $\beta = \frac{n}{\eta}$. Substituting Equation \ref{eq:boost} into Equation \ref{eq:updaterule}, and assuming convergence of benign clients, leads to the following approximation:

\begin{equation*}
    {V_\mathcal{G}}^{t+1} \approx {V_\mathcal{G}}^{t} + \frac{\eta}{n} \frac{n}{\eta} ({V}_{adv}^t - {V_\mathcal{G}}^{t}) = {V}_{adv}^t.
\end{equation*}

This result demonstrates that, under the model replacement strategy, the global model can be effectively overridden by the adversarial update, thereby granting the malicious client disproportionate influence over the training process.

\subsection{Threat Model}\label{sec:threat}
\rewrite{To evaluate the resilience of the RFC framework, we formalize the adversary model within a decentralized learning environment. We assume a system consisting of $n$ total clients $\mathcal{C} = \{C_1, \dots, C_n\}$, partitioned into $M$ disjoint mining pools $\{\mathcal{P}_1, \dots, \mathcal{P}_M\}$. We consider a Byzantine adversary $\mathcal{A}$ with the following properties:}
\begin{itemize}
\item \rewrite{\textbf{Control and Collusion:} The adversary controls a subset of clients $\mathcal{C}_{adv} \subset \mathcal{C}$. Malicious clients within the same pool $\mathcal{P}_i$ are assumed to collude to maximize the impact of their corrupted updates $\hat{V}_{adv}^t$ (i.e., they cooperate to maximize the attack impact).}
\item \rewrite{\textbf{White-box Knowledge:} Following the worst-case security analysis, $\mathcal{A}$ has full knowledge of the global model parameters $G^t$, the learning rate $\eta$, and the specific robust aggregation rule $\mathcal{R}$ employed by the pool miner.}
\item \rewrite{\textbf{Model Replacement:} The adversary can employ model replacement strategies by scaling malicious updates by a factor $\beta = \frac{n}{\eta}$ to override the global model state.}
\end{itemize}

\subsection{Baselines}\label{sec:baselines}
To empirically assess the efficacy of the RFC schema, we define a clear set of baseline aggregation rules for comparison. Specifically, we consider the canonical FedAvg together with three widely used robust \rewrite{aggregators}: Krum, Bulyan, and the Geometric Median (GeoMed). Throughout, let $\{V_1,\ldots,V_n\}$ denote the set of client updates.

\begin{itemize}
    \item \textbf{FedAvg}~\cite{bib:mcmahan16communicationefficient}. The original and most widely deployed aggregation rule in FL. It computes the global update as the arithmetic mean of all received client updates:
    \begin{equation*}
        FA(V_1, \ldots, V_n) = \frac{1}{n}\sum_{i=1}^{n} V_i .
    \end{equation*}
    While simple and effective under benign conditions, FedAvg is not robust and is therefore susceptible to a variety of adversarial manipulations.

    \item \textbf{Krum}~\cite{krum}. Krum selects the single update in $\{V_1,\ldots,V_n\}$ that minimizes its total (squared) Euclidean distance to a designated subset of other updates. Formally,
    \begin{equation*}
        KR(V_1, \ldots, V_n) = \underset{i \in \{1, \ldots, n\}}{\operatorname{argmin}} \ s(i),
    \end{equation*}
    where the score
    \begin{equation*}
        s(i) = \sum_{i \to j} \| V_i - V_j \|^2
    \end{equation*}
    accumulates distances over the selected neighbor subset.

    \item \textbf{Bulyan}~\cite{bulyan}. Bulyan builds on Krum by first computing Krum scores $s(i)$ for all updates and then selecting a subset $\mathcal{V}$ of $m$ updates with the smallest scores. The final aggregate is the average over the selected subset:
    \begin{equation*}
        B(V_1, \ldots, V_n) = \frac{1}{|\mathcal{V}|}\sum_{V \in \mathcal{V}} V ,
    \end{equation*}
    where $\mathcal{V} \subseteq \{V_1, \ldots, V_n\}$ contains the $m$ lowest-score updates. Following common practice, we set $m=5$ in our experiments.

    \item \textbf{GeoMed}~\cite{bulyan,rousseeuw85}. The Geometric Median aggregator selects the update
    \begin{equation*}
        Geo(V_1, \ldots, V_n) = \underset{V_j \in \{V_1, \ldots, V_n\}}{\operatorname{argmin}} \ \sum_{i =1}^n \| V_i - V_j \|^2 ,
    \end{equation*}
    i.e., the candidate that is most centrally located with respect to the others in Euclidean distance.
\end{itemize}


\subsection{Evaluation metrics}\label{sec:metrics}
In order to provide a holistic analysis for the proposal, we provide a range of different metrics. For both accuracy and loss, we provide the following variants of these metrics.

\begin{itemize}
    \item \textbf{Final.} It \rewrite{consists of} the final value of this metric. That is, the value in the last learning round. It is a common metric for model evaluation.
    \item \textbf{Max/Min.} It \rewrite{consists of} the maximum (minimum) value for the accuracy (loss), which \rewrite{represents} the best value obtained during the whole training. In case we are able to use an evaluation dataset, this would be the result of the selected model.
    \item \textbf{Avg last 10.} It \rewrite{consists of} the mean value of the given metric over the last 10 learning rounds. It \rewrite{allows} us to check for resilience over training, since a model may present \rewrite{significant fluctuations in the metrics}, which may not be reflected on the previously explained metrics.
\end{itemize}

In the case of a backdoor attack, we present the same metrics for the malicious secondary task, allowing us to also analyze the success of the attack.

\subsection{Implementation details}\label{sec:details}
To ensure reproducibility, we provide the code used to run all experiments~\footnote{\url{https://github.com/ari-dasci/S-xfc}}. The code is written using the FLEXible FL framework~\cite{herrera2024flex}~\footnote{\url{https://github.com/FLEXible-FL/FLEXible}} and its companion libraries \texttt{flex-clash}~\footnote{\url{https://github.com/FLEXible-FL/flex-clash}} to simulate attacks on the FL scheme and access basic federated aggregation rules, and \texttt{flex-block}~\footnote{\url{https://github.com/FLEXible-FL/flex-block}} to implement and simulate the Blockchain architectures employed; also PyTorch~\cite{pytorch} has been used for implementing the models.

\section{Experimental results and analysis}\label{sec:analysis}
\rewrite{This section presents the experimental results and corresponding analysis to validate the proposed methodology. In order to provide an exhaustive and clear analysis of the framework, we first pose a set of research questions that will help us to extract insight from results. The questions are as follows:}
\begin{itemize}
    \item \rewrite{\textbf{RQ1 (Performance baseline).} How does the RFC framework affect model convergence and final performance in benign (non-adversarial) environments compared to standard FL and PoFL?}
    \item \rewrite{\textbf{RQ2 (Fault isolation).} To what extent does RFC's pool redundancy effectively isolate and mitigate single-point adversarial injections (Byzantine and Backdoor)?}
    \item \rewrite{\textbf{RQ3 (Cumulative Resilience).} Can the hierarchical combination of intra-pool robust aggregation and blockchain-level consensus withstand a coordinated multi-pool attack where the majority of clusters are compromised?}
    \item \rewrite{\textbf{RQ4 (Metric Adaptability).} Does the decoupling of the evaluation metric from differentiability allow for better mitigation compared to rigid accuracy-based consensus?}
\end{itemize}

\rewrite{In order to answer \rewrite{these} questions, we evaluate the baselines in both the standard \rewrite{client-server} case and its \rewrite{blockchain-based} RFC counterpart \rewrite{(i.e., we fix the aggregator hyperparameter to each of those already introduced)}: Krum Federated Chain (K-RFC), Bulyan Federated Chain (B-RFC), and GeoMed Federated Chain (G-RFC). The RFC variant of FedAvg recovers the original PoFL schema, extended here with the capacity to optimize objective functions beyond raw model accuracy, which we will also call PoFL for the sake of conciseness.}

\noindent\textbf{Evaluation scenarios.} \rewrite{Each architecture is assessed under two high-level conditions:}
\begin{itemize}
    \item \textbf{No attack.} \rewrite{All clients behave benignly. Shown in Section~\ref{sec:no_attack}}
    \item \textbf{Under attack.} \rewrite{Adversarial clients are present during training, with two subcases:}
    \begin{itemize}
        \item \textbf{One miner compromised.} \rewrite{At any given time, exactly one miner includes at least one adversarial client in its training pool. Shown in Section~\ref{sec:one_client}.}
        \item \textbf{All miners compromised.} \rewrite{Every miner includes at least one adversarial client in its training pool. In the \rewrite{client-server} architecture, this case is equivalent to the previous subcase. Shown in Section~\ref{sec:mult_clients}.}
    \end{itemize}
\end{itemize}

\rewrite{We do not \rewrite{guarantee} that for a given pool $P_i$, the requirements of a given aggregation rule are satisfied. For example, Krum or Bulyan \rewrite{need} the proportion of adversarial clients within a pool \rewrite{to be} less than the specified hyperparameter $f$. However, we \rewrite{guarantee} that \rewrite{these conditions are met} at least in one pool.}


\subsection{No Attack Scenario}\label{sec:no_attack}
\begin{figure}
    \centering
    \begin{subfigure}[h]{0.45\textwidth}
        \centering
        \includegraphics[width=\textwidth]{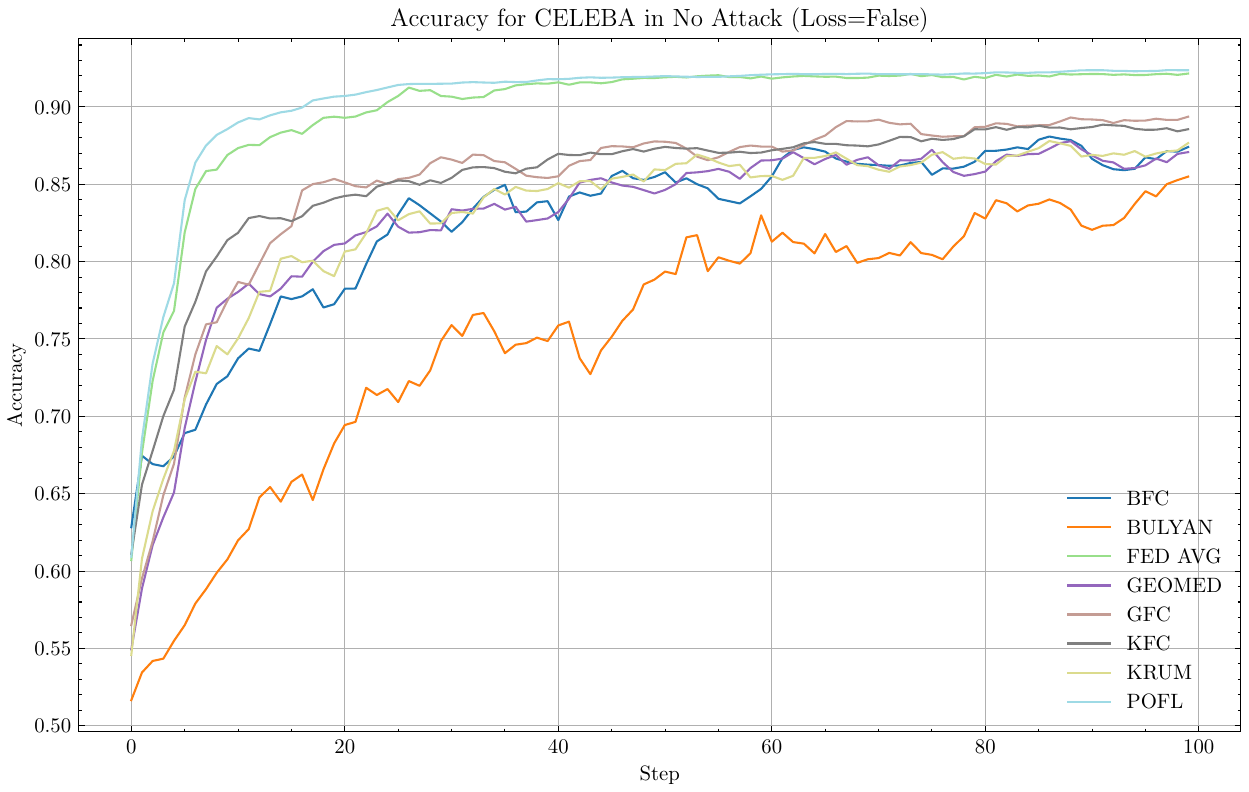}
        \caption{Accuracy over the rounds on CelebA-S dataset. Using accuracy as evaluation metric.}
        \label{fig:no_attack_celeba_acc}
    \end{subfigure}
    \hfill
    \begin{subfigure}[h]{0.45\textwidth}
        \centering
        \includegraphics[width=\textwidth]{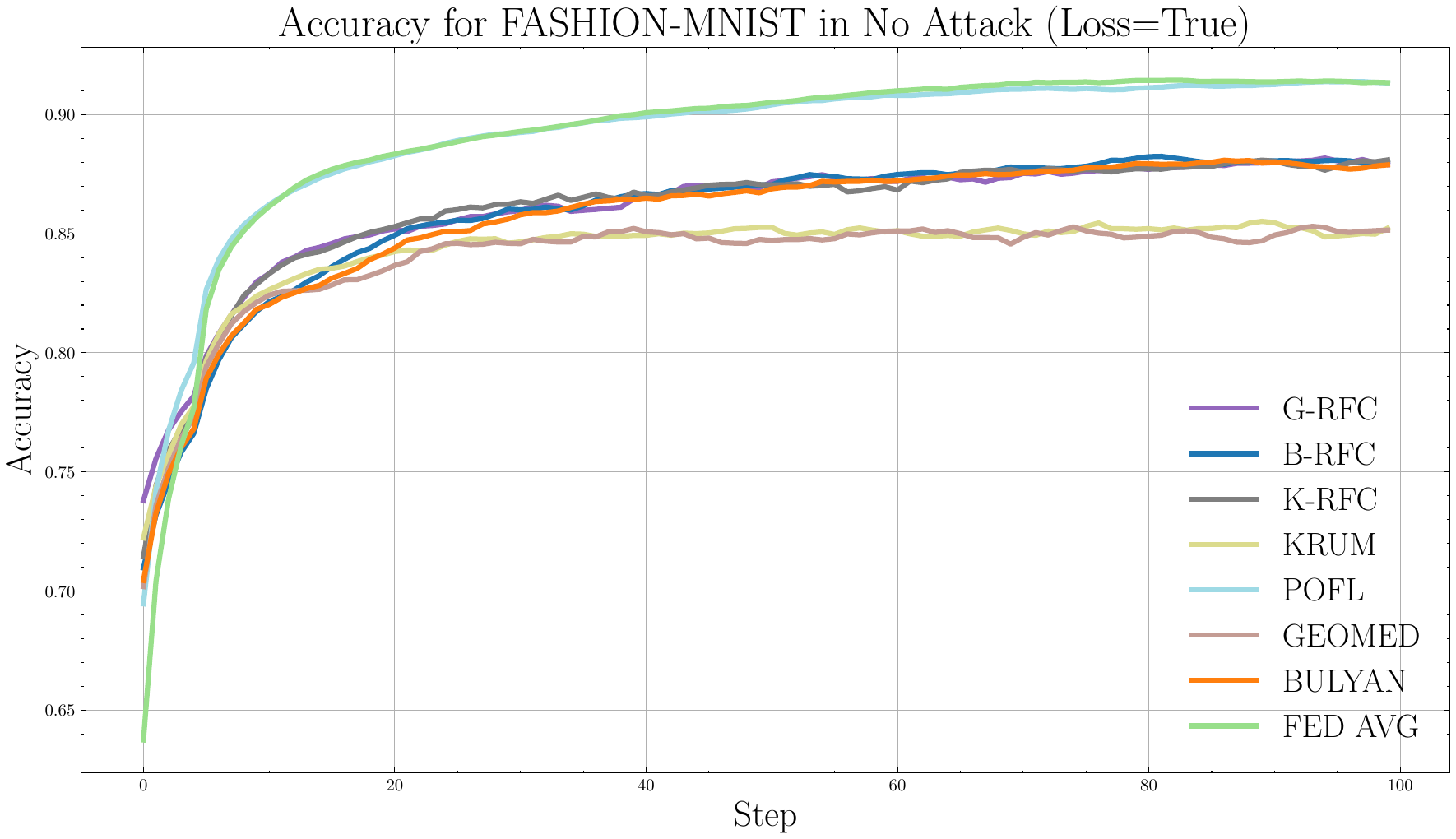}
        \caption{Accuracy over the rounds on Fashion-MNIST dataset. Using loss as evaluation metric.}
        \label{fig:no_attack_fashion_loss}
    \end{subfigure}
\caption{Performance of the proposal under the no attack scenario. We see how \paco{RFC} variants are superior to their baselines both in Figures \ref{fig:no_attack_celeba_acc} and \ref{fig:no_attack_fashion_loss}. Non robust aggregation operators obtain the better accuracy, which is an expected result.}
\label{fig:no_attack}
\end{figure}

\rewrite{In this scenario, we consider that all clients behave benignly, that is, no adversarial client is present during the training procedure. By considering this scenario, we aim to provide \rewrite{answers} to both \textbf{RQ1} and \textbf{RQ4}. This is important since in a real production system, the presence of adversarial attacks is unknown and \rewrite{thus}, a huge degradation of the system performance may concern potential adopters of robust methods. All the evaluation metrics and relevant evolution of those are presented in Table~\ref{tab:no-attack} and Figure~\ref{fig:no_attack}.}

\rewrite{For \textbf{RQ1}, we see how the proposed system demonstrates consistent performance across all datasets and evaluation metrics. While FedAvg and PoFl achieve the best performance, which is an expected outcome given the general performance degradation associated with robust aggregation techniques~\cite{krum}, the RFC framework consistently outperforms the original aggregators in every variant. This suggests that even without malicious clients, the RFC architecture provides a performance improvement. When considered alongside its other benefits, such as enhanced scalability and the elimination of a single point of failure, this makes the proposed architecture suitable for application in this scenario. Furthermore, regarding \textbf{RQ4}, in this specific scenario, optimizing for either accuracy or loss yields similar metrics, which suggests the training procedure was successful across all pools.}

\begin{table}[!ht]
\centering
\tiny
\caption{Results obtained in the no attack scenario, showing both results from the loss and accuracy based.}
\label{tab:no-attack}
\begin{tabular}{l|l|rrr|rrr|rrr|rrr}
	\toprule
\multicolumn{2}{c|}{} & \multicolumn{6}{c|}{Loss based} & \multicolumn{6}{c}{Accuracy based} \\
\cmidrule(lr){3-8} \cmidrule(lr){9-14}
&            &         \multicolumn{3}{c|}{Accuracy}          &            \multicolumn{3}{c|}{Loss}            &         \multicolumn{3}{c|}{Accuracy}         &            \multicolumn{3}{c}{Loss}             \\ 
	Dataset                                                                                                       & Aggregator &         Final &          Mean &            Max &          Final &          Mean &            Min &         Final &          Mean &           Max &          Final &           Mean &           Min \\
	                                                                                                              &            &               &       Last 10 &                &                &       Last 10 &                &               &       Last 10 &               &                &        Last 10 &               \\ \midrule
	\multirow{9}{*}{Celeba-S}                                                                                     & \paco{B-RFC}        &          0.91 &          0.92 &           0.92 & \textbf{ 0.26} & \textbf{0.26} &           0.25 &           0.88 &           0.87 &           0.89 &           0.66 &           0.60 &           0.45 \\
	                                                                                                              & Bulyan     &          0.92 &          0.92 &           0.92 &           0.27 &          0.27 &           0.27 &           0.87 &           0.85 &           0.87 &           0.55 &           0.57 &           0.45 \\
	                                                                                                              & \paco{G-RFC}        &          0.91 &          0.92 &           0.92 &           0.27 &          0.27 &           0.25 &           0.89 &           0.89 &           0.90 &           0.59 &           0.46 &  \textbf{0.30} \\
	                                                                                                              & GeoMed     &          0.91 &          0.92 &           0.92 &           0.29 &          0.29 &           0.27 &           0.86 &           0.87 &           0.89 &           0.43 &           0.47 &           0.40 \\
	                                                                                                              & \paco{K-RFC}        &          0.91 &          0.92 &           0.92 &  \textbf{0.26} & \textbf{0.26} &           0.25 &           0.89 &           0.89 &           0.90 &           0.40 &           0.44 &           0.38 \\
	                                                                                                              & Krum       &          0.92 &          0.92 &           0.92 &           0.27 &          0.27 &           0.26 &           0.88 &           0.87 &           0.89 &           0.44 &           0.44 &           0.36 \\
	                                                                                                              & PoFL       & \textbf{0.93} & \textbf{0.93} &  \textbf{0.93} &  \textbf{0.26} & \textbf{0.26} &  \textbf{0.23} &  \textbf{0.92} &  \textbf{0.92} &           0.92 &           0.26 &           0.26 &  \textbf{0.23} \\ 
	                                                                                                              & FedAvg     & \textbf{0.93} & \textbf{0.93} &  \textbf{0.93} &           0.31 &          0.30 &  \textbf{0.23} &  \textbf{0.92} &  \textbf{0.92} &  \textbf{0.93} &  \textbf{0.25} &  \textbf{0.25} &  \textbf{0.22} \\ \midrule
	\multirow{9}{*}{CIFAR-10}                                                                                     & \paco{B-RFC}        &          0.80 &          0.80 &           0.81 &           0.67 &          0.69 &           0.65 &          0.80 &          0.80 &          0.81 &           0.74 &           1.06 &          0.65 \\
	                                                                                                              & Bulyan     &          0.81 &          0.81 &           0.82 &           0.67 &          0.66 &           0.62 &          0.80 &          0.80 &          0.80 &           0.68 &           0.68 &          0.67 \\
	                                                                                                              & \paco{G-RFC}        &          0.84 &          0.84 &           0.84 &           0.52 &          0.55 &           0.52 &          0.84 &          0.83 &          0.84 &           0.56 &           0.58 &          0.54 \\
	                                                                                                              & GeoMed     &          0.84 &          0.84 &           0.84 &           0.55 &          0.56 &           0.54 &          0.84 &          0.83 &          0.84 &           0.57 &           0.57 &          0.54 \\
	                                                                                                              & \paco{K-RFC}        &          0.84 &          0.84 &           0.84 &           0.53 &          0.56 &           0.53 &          0.83 &          0.83 &          0.84 &           0.60 &           0.59 &          0.55 \\
	                                                                                                              & Krum       &          0.84 &          0.83 &           0.84 &           0.54 &          0.57 &           0.54 &          0.83 &          0.83 &          0.84 &           0.60 &           0.58 &          0.55 \\
	                                                                                                              & PoFL       &          0.96 &          0.96 &           0.96 &  \textbf{0.18} &          0.18 &           0.15 &          0.96 &          0.96 &          0.96 &           0.19 &           0.19 &          0.15 \\ 
	                                                                                                              & FedAvg     & \textbf{0.97} & \textbf{0.97} & \textbf{ 0.97} &  \textbf{0.17} & \textbf{0.17} &  \textbf{0.13} & \textbf{0.97} & \textbf{0.97} & \textbf{0.97} &  \textbf{0.18} &  \textbf{0.17} & \textbf{0.13} \\ \midrule
	\multirow{9}{*}{Fashion MNIST}                                                                                & \paco{B-RFC}        &          0.88 &          0.88 &           0.88 &           0.45 &          0.43 &           0.39 &          0.88 &          0.88 &          0.88 &           0.45 &           0.45 &          0.42 \\
	                                                                                                              & Bulyan     &          0.88 &          0.88 &           0.88 &           0.46 &          0.49 &           0.43 &          0.88 &          0.88 &          0.88 &           0.46 &           0.45 &          0.43 \\
	                                                                                                              & \paco{G-RFC}        &          0.88 &          0.88 &           0.89 &           0.55 &          0.58 &           0.46 &          0.87 &          0.87 &          0.87 &           1.21 &           0.96 &          0.49 \\
	                                                                                                              & GeoMed     &          0.85 &          0.85 &           0.86 &           1.81 &          1.56 &           0.58 &          0.85 &          0.85 &          0.86 &           1.83 &           1.67 &          0.57 \\
	                                                                                                              & \paco{K-RFC}        &          0.88 &          0.88 &           0.88 &           0.62 &          0.58 &           0.45 &          0.88 &          0.87 &          0.88 &           0.69 &           0.81 &          0.50 \\
	                                                                                                              & Krum       &          0.86 &          0.85 &           0.86 &           1.70 &          1.39 &           0.56 &          0.85 &          0.85 &          0.86 &           1.91 &           1.81 &          0.56 \\
	                                                                                                              & PoFL       & \textbf{0.91} & \textbf{0.91} &           0.91 &           0.31 &          0.31 &           0.31 & \textbf{0.91} & \textbf{0.91} & \textbf{0.91} &           0.34 &           0.33 &          0.32 \\ 
	                                                                                                              & FedAvg     & \textbf{0.91} & \textbf{0.91} &  \textbf{0.92} &  \textbf{0.30} & \textbf{0.30} & \textbf{ 0.30} & \textbf{0.91} & \textbf{0.91} & \textbf{0.91} &  \textbf{0.31} &  \textbf{0.30} & \textbf{0.30} \\ \midrule
	\multirow{9}{*}{EMNIST}                                                                                       & \paco{B-RFC}        &          0.98 &          0.98 &           0.98 &           0.11 &          0.10 &           0.08 &          0.98 &          0.98 &          0.98 &           0.12 &           0.11 &          0.09 \\
	                                                                                                              & Bulyan     &          0.98 &          0.98 &           0.98 &           0.11 &          0.11 &           0.09 &          0.97 &          0.98 &          0.98 &           0.10 &           0.10 &          0.10 \\
	                                                                                                              & \paco{G-RFC}        &          0.98 &          0.98 &           0.98 &           0.20 &          0.21 &           0.11 &          0.98 &          0.97 &          0.98 &           0.32 &           0.31 &          0.13 \\
	                                                                                                              & GeoMed     &          0.97 &          0.97 &           0.97 &           0.49 &          0.47 &           0.20 &          0.97 &          0.97 &          0.97 &           0.48 &           0.48 &          0.19 \\
	                                                                                                              & \paco{K-RFC}        &          0.97 &          0.97 &           0.98 &           0.14 &          0.16 &           0.11 &          0.97 &          0.97 &          0.98 &           0.36 &           0.37 &          0.14 \\
	                                                                                                              & Krum       &          0.96 &          0.96 &           0.97 &           0.51 &          0.47 &           0.21 &          0.97 &          0.97 &          0.97 &           0.41 &           0.42 &          0.18 \\
	                                                                                                              & PoFL       & \textbf{0.99} & \textbf{0.99} &  \textbf{0.99} &  \textbf{0.05} &          0.05 &  \textbf{0.04} & \textbf{0.99} & \textbf{0.99} & \textbf{0.99} &           0.05 &           0.05 &          0.05 \\ 
	                                                                                                              & FedAvg     & \textbf{0.99} & \textbf{0.99} & \textbf{ 0.99} &  \textbf{0.05} & \textbf{0.04} &  \textbf{0.04} & \textbf{0.99} & \textbf{0.99} & \textbf{0.99} & \textbf{ 0.04} & \textbf{ 0.04} & \textbf{0.04} \\ \bottomrule
\end{tabular}
\end{table}

\subsection{One Pool Attack Scenario}\label{sec:one_client}
\rewrite{We move into the scenario where adversarial clients are present during the training of the model. First, we will study the scenario when only one pool is compromised, being this the most basic threat model that we consider. We study two kinds of attacks, a byzantine attack and a backdoor attack. In this scenario we will try to answer \textbf{RQ2} and \textbf{RQ4}.}

\rewrite{For \textbf{RQ2}, Table \ref{tab:one-client-labelflip} presents the results of a Byzantine attack, specifically a labelflip attack. These results are largely consistent with the \rewrite{no-attack} scenario, which is expected since the majority of the aggregation methods employed are designed to be resilient in such a situation and \rewrite{thus}, showing that RFC is resilient to adversarial attacks. However, notable differences emerge upon closer examination.}

A critical observation is the degraded performance of FedAvg, which aligns with our hypothesis that this aggregation technique lacks inherent robustness. This is in stark contrast to PoFL, which maintains stable performance, \rewrite{thus \rewrite{furthering} our assumption that its architecture is inherently resilient to this type of attack and fully answering \textbf{RQ2}, which shows that the answer to this question does not depend on the selected aggregator.} In fact, for the Celeba-S dataset, the FedAvg model's performance degrades significantly, dropping to around 46\% accuracy. For the CIFAR-10 dataset, this degradation is even more severe, resulting in NaN loss values. We hypothesize that this is caused by the failure of the batch normalization layers in the EfficientNet-B0 model under the byzantine attack.

\begin{figure}
    \centering
    \begin{subfigure}[h]{0.45\textwidth}
        \centering
        \includegraphics[width=\textwidth]{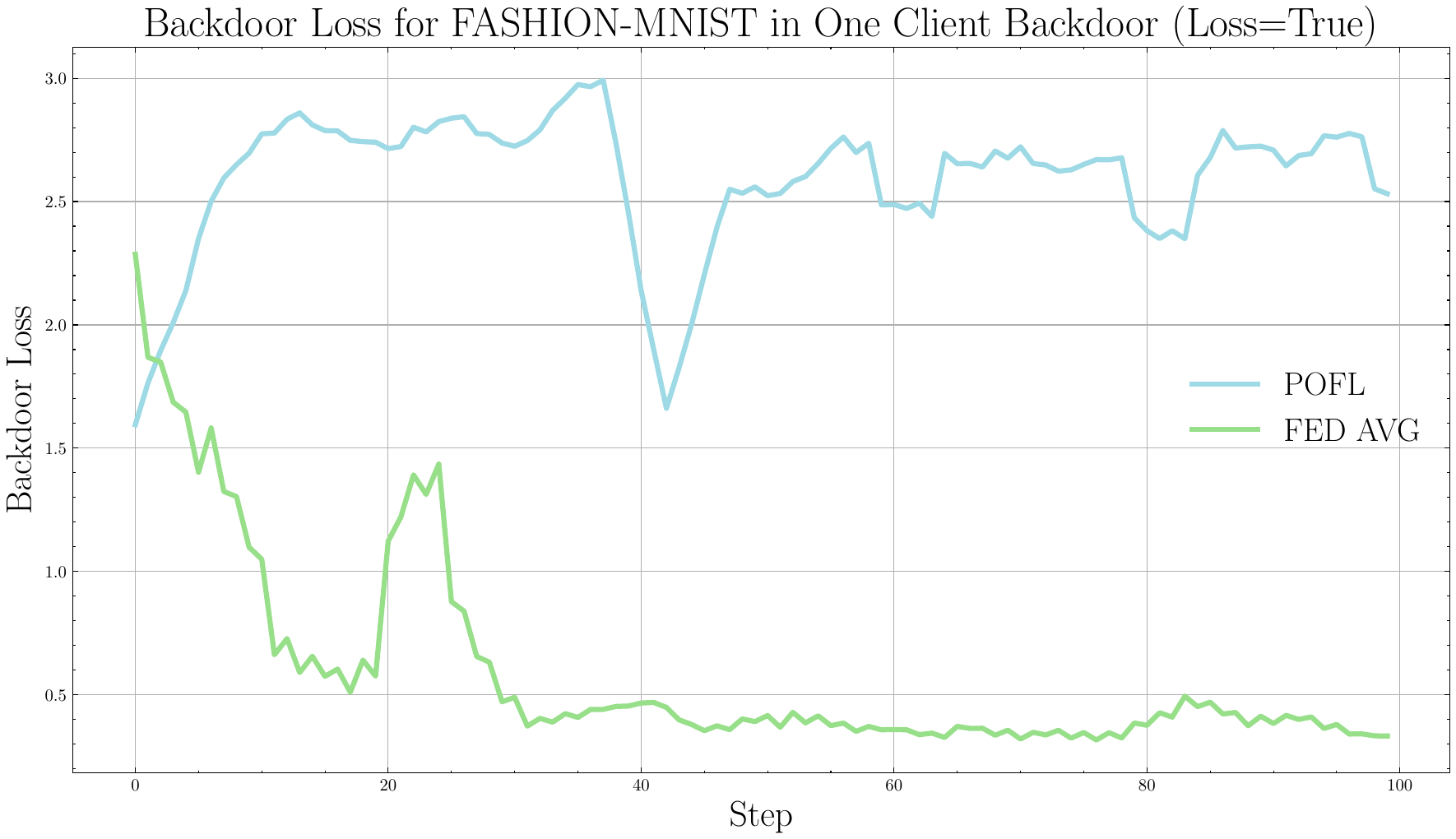}
        \caption{Backdoor loss over the rounds on Fashion-MNIST dataset. Using loss as evaluation metric.}
        \label{fig:one_backdoor_fashion_loss}
    \end{subfigure}
    \hfill
    \begin{subfigure}[h]{0.45\textwidth}
        \centering
        \includegraphics[width=\textwidth]{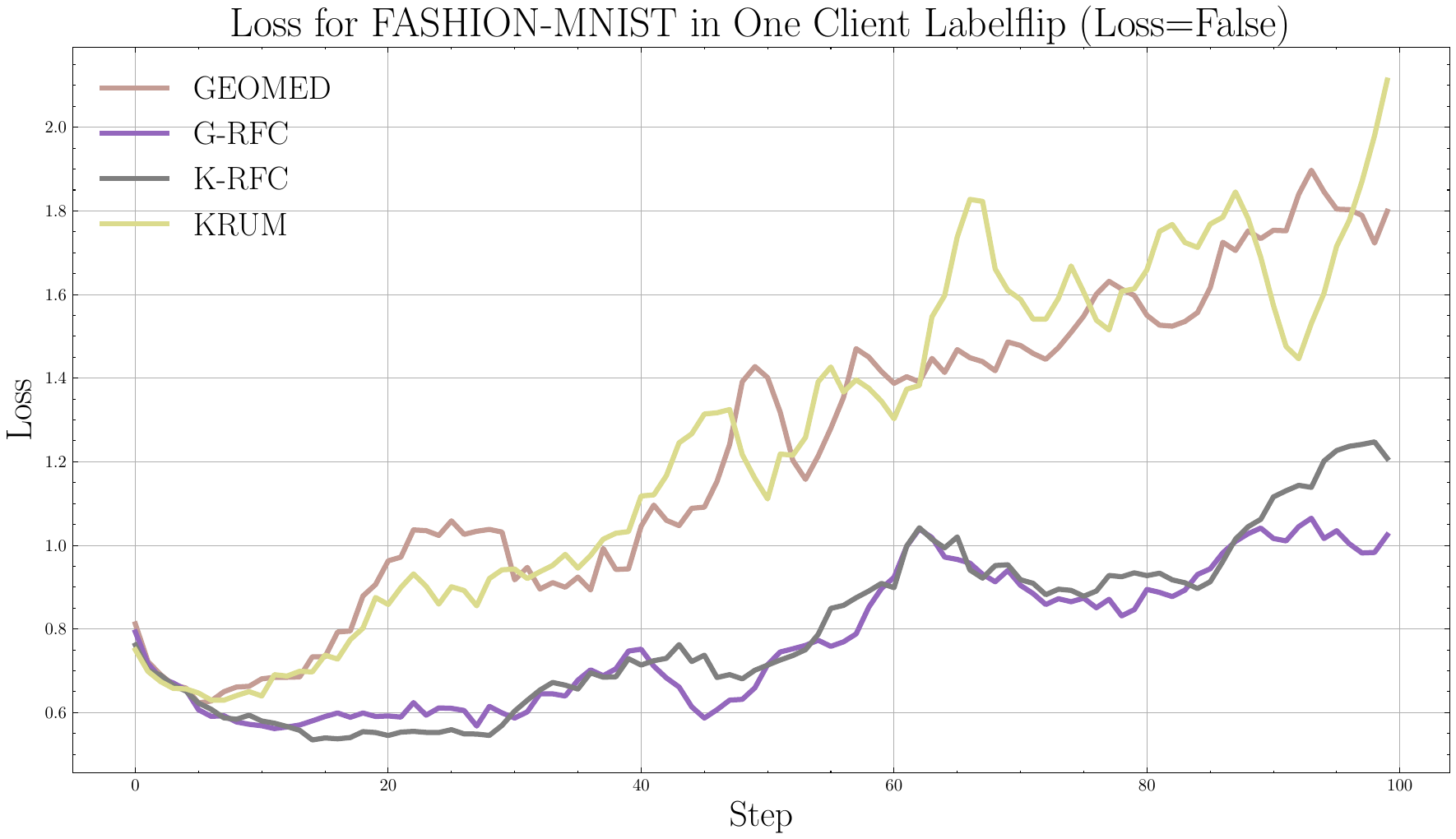}
        \caption{Labelflip loss over the rounds on Fashion-MNIST dataset. Using accuracy as evaluation metric.}
        \label{fig:one_labelflip_fashion_loss}
    \end{subfigure}

    \begin{subfigure}[h]{0.45\textwidth}
        \centering
        \includegraphics[width=\textwidth]{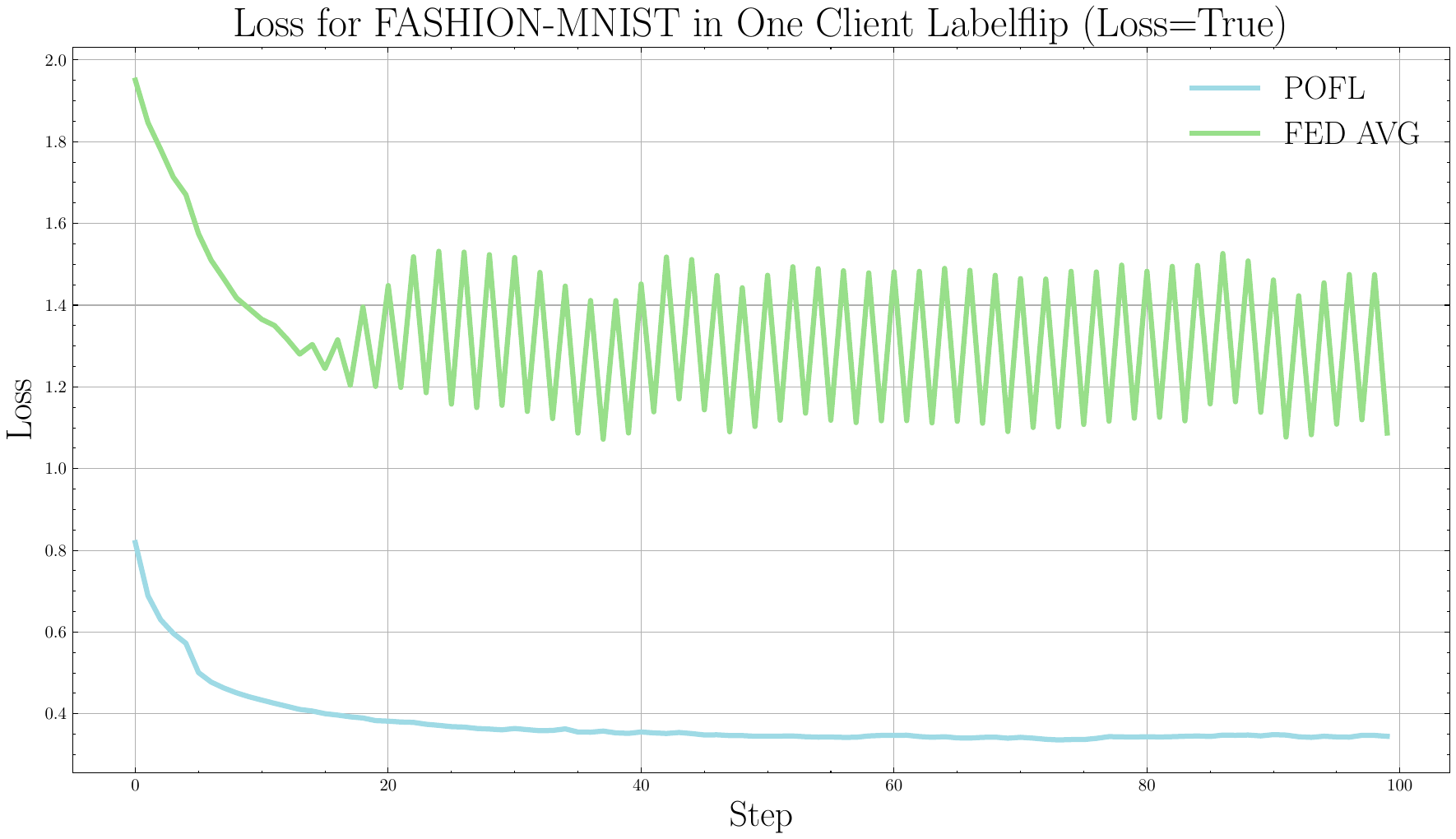}
        \caption{Labelflip loss over the rounds on Fashion-MNIST dataset. Using loss as evaluation metric.}
        \label{fig:one_labelflip_fashion_loss_pofl}
    \end{subfigure}

\caption{Performance of the proposal under the one pool attack scenario. Figure \ref{fig:one_backdoor_fashion_loss} and \ref{fig:one_labelflip_fashion_loss_pofl} shows how the PoFL framework effectively mitigates both backdoor and labelflip attacks while FedAvg fails to do so. Figure \ref{fig:one_labelflip_fashion_loss} illustrates the overfitting issue observed with Krum and GeoMed in the labelflip attack, which is mitigated by their \paco{RFC} variants.}
\label{fig:one_attack}
\end{figure}

\rewrite{Moving \rewrite{to} \textbf{RQ4}, we observed that robust aggregation rules like Krum and GeoMed tend to overfit in this scenario. For instance, on the Fashion MNIST dataset, these two aggregators show decent accuracy but exhibit a significantly higher loss compared to other methods. Their RFC variants, specifically K-RFC and G-RFC, effectively mitigate this overfitting, as evidenced by their lower and more stable loss values. These findings collectively suggest that RFC not only serves as a valid defense but also offers an improved and more stable solution in this adversarial context. These insights are illustrated in Figures~\ref{fig:one_labelflip_fashion_loss} and ~\ref{fig:one_labelflip_fashion_loss_pofl}.}

\begin{table}[!ht]
\centering
\tiny
\caption{One-client scenario with Labelflip attack: loss-based and accuracy-based evaluations.}
\label{tab:one-client-labelflip}
\begin{tabular}{l|l|rrr|rrr|rrr|rrr}
	\toprule
	           \multicolumn{2}{c|}{}            &                                 \multicolumn{6}{c|}{Loss based}                                  &                               \multicolumn{6}{c}{Accuracy based}                                \\ \cmidrule
	(lr){3-8} \cmidrule(lr){9-14}  &            &          \multicolumn{3}{c|}{Accuracy}          &           \multicolumn{3}{c|}{Loss}            &         \multicolumn{3}{c|}{Accuracy}          &            \multicolumn{3}{c}{Loss}            \\
	                        &  &          Final &          Mean &            Max &         Final &          Mean &            Min &         Final &          Mean &            Max &         Final &           Mean &           Min \\
	Dataset                        & Aggregator &                &       Last 10 &                &               &       Last 10 &                &               &       Last 10 &                &               &        Last 10 &               \\ \midrule
	\multirow{9}{*}{Celeba-S}      & \paco{B-RFC}        &           0.89 &          0.88 &           0.89 &          0.72 &          0.56 &           0.36 &           0.88 &           0.88 &           0.90 &           0.41 &           0.60 &           0.34 \\
	                               & Bulyan     &           0.87 &          0.80 &           0.88 &          0.41 &          0.68 &           0.38 &           0.85 &           0.78 &           0.89 &           0.43 &           0.85 &           0.34 \\
	                               & \paco{G-RFC}        &           0.89 &          0.88 &           0.89 &          0.28 &          0.41 &           0.28 &             -- &             -- &             -- &             -- &             -- &             -- \\
	                               & GeoMed     &           0.88 &          0.85 &           0.89 &          0.38 &          0.54 &           0.36 &           0.88 &           0.88 &           0.90 &           0.51 &           0.46 &           0.36 \\
	                               & \paco{K-RFC}        &           0.86 &          0.88 &           0.90 &          0.36 &          0.37 &           0.31 &           0.86 &           0.88 &           0.90 &           0.45 &           0.46 &           0.35 \\
	                               & Krum       &           0.89 &          0.87 &           0.89 &          0.38 &          0.46 &           0.32 &           0.85 &           0.87 &           0.89 &           0.59 &           0.52 &           0.31 \\
	                               & PoFL       &  \textbf{0.92} & \textbf{0.92} &  \textbf{0.92} & \textbf{0.26} & \textbf{0.24} &  \textbf{0.20} &  \textbf{0.92} &  \textbf{0.92} &  \textbf{0.93} &  \textbf{0.27} &  \textbf{0.29} &  \textbf{0.20} \\ 
	                               & FedAvg     &           0.46 &          0.47 &           0.55 &          1.19 &          1.39 &           0.91 &           0.44 &           0.42 &           0.56 &           1.45 &           1.60 &           0.94 \\ \midrule
	\multirow{8}{*}{CIFAR-10}      & \paco{B-RFC}        &           0.79 &          0.80 &           0.80 &          0.70 &          0.69 &           0.65 &          0.80 &          0.80 &           0.81 &          0.71 &           0.87 &          0.65 \\
	                               & Bulyan     &           0.83 &          0.83 &           0.84 &          0.57 &          0.57 &           0.54 &          0.82 &          0.83 &           0.84 &          0.64 &           0.61 &          0.56 \\
	                               & \paco{G-RFC}        &           0.84 &          0.84 &           0.85 &          0.56 &          0.55 &           0.52 &          0.84 &          0.84 &           0.85 &          0.54 &           0.56 &          0.52 \\
	                               & GeoMed     &           0.84 &          0.83 &           0.84 &          0.57 &          0.58 &           0.55 &          0.84 &          0.83 &           0.84 &          0.54 &           0.58 &          0.54 \\
	                               & \paco{K-RFC}        &           0.84 &          0.84 &           0.85 &          0.54 &          0.55 &           0.54 &          0.84 &          0.84 &           0.85 &          0.58 &           0.57 &          0.53 \\
	                               & Krum       &           0.82 &          0.83 &           0.84 &          0.64 &          0.59 &           0.56 &          0.84 &          0.83 &           0.84 &          0.59 &           0.58 &          0.55 \\
	                               & PoFL       &  \textbf{0.96} & \textbf{0.96} &  \textbf{0.96} & \textbf{0.23} & \textbf{0.23} &  \textbf{0.16} & \textbf{0.96} & \textbf{0.96} & \textbf{ 0.96} & \textbf{0.24} &  \textbf{0.22} &          0.17 \\ 
	                               & FedAvg     &           0.10 &          0.12 &           0.20 &           NaN &          2.47 &           2.24 &          0.10 &          0.13 &           0.39 &           NaN &           2.12 &          2.12 \\ \midrule
	\multirow{8}{*}{Fashion MNIST} & \paco{B-RFC}        &           0.88 &          0.88 &           0.88 &          0.42 &          0.44 &           0.42 &          0.88 &          0.88 &           0.88 &          0.48 &           0.46 &          0.43 \\
	                               & Bulyan     &           0.88 &          0.89 &           0.89 &          0.49 &          0.45 &           0.40 &          0.88 &          0.88 &           0.89 &          0.46 &           0.46 &          0.43 \\
	                               & \paco{G-RFC}        &           0.88 &          0.88 &           0.88 &          0.60 &          0.58 &           0.47 &          0.87 &          0.87 &           0.87 &          1.11 &           1.02 &          0.49 \\
	                               & GeoMed     &           0.85 &          0.85 &           0.86 &          1.84 &          1.92 &           0.59 &          0.85 &          0.85 &           0.85 &          1.79 &           1.82 &          0.59 \\
	                               & \paco{K-RFC}        &           0.87 &          0.87 &           0.88 &          0.70 &          0.68 &           0.46 &          0.86 &          0.86 &           0.87 &          1.08 &           1.20 &          0.52 \\
	                               & Krum       &           0.85 &          0.85 &           0.86 &          1.82 &          2.14 &           0.56 &          0.85 &          0.85 &           0.85 &          2.35 &           1.86 &          0.56 \\
	                               & PoFL       &  \textbf{0.91} & \textbf{0.91} &  \textbf{0.91} & \textbf{0.34} & \textbf{0.34} & \textbf{ 0.33} & \textbf{0.91} & \textbf{0.91} &  \textbf{0.91} & \textbf{0.37} & \textbf{ 0.37} & \textbf{0.33} \\ 
	                               & FedAvg     &           0.88 &          0.51 &           0.89 &          0.37 &          1.33 &           0.34 &          0.88 &          0.67 &           0.89 &          0.35 &           1.27 & \textbf{0.33} \\ \midrule
	\multirow{8}{*}{EMNIST}        & \paco{B-RFC}        &           0.97 &          0.97 &           0.98 &          0.14 &          0.11 &           0.08 &          0.98 &          0.97 &           0.98 &          0.11 &           0.12 &          0.10 \\
	                               & Bulyan     &           0.98 &          0.98 &           0.98 &          0.11 &          0.10 &           0.08 &          0.98 &          0.98 &           0.98 &          0.11 &           0.10 &          0.08 \\
	                               & \paco{G-RFC}        &           0.97 &          0.97 &           0.98 &          0.25 &          0.23 &           0.13 &          0.97 &          0.97 &           0.97 &          0.35 &           0.34 &          0.15 \\
	                               & GeoMed     &           0.96 &          0.96 &           0.97 &          0.65 &          0.54 &           0.20 &          0.97 &          0.97 &           0.97 &          0.49 &           0.49 &          0.21 \\
	                               & \paco{K-RFC}        &           0.97 &          0.97 &           0.98 &          0.19 &          0.19 &           0.12 &          0.97 &          0.97 &           0.97 &          0.38 &           0.40 &          0.15 \\
	                               & Krum       &           0.96 &          0.96 &           0.96 &          0.69 &          0.69 &           0.26 &          0.96 &          0.96 &           0.96 &          0.74 &           0.69 &          0.24 \\
	                               & PoFL       & \textbf{ 0.99} & \textbf{0.99} &  \textbf{0.99} & \textbf{0.05} & \textbf{0.05} &  \textbf{0.05} & \textbf{0.99} & \textbf{0.99} &  \textbf{0.99} & \textbf{0.06} &  \textbf{0.05} & \textbf{0.05} \\ 
	                               & FedAvg     &           0.11 &          0.11 &           0.98 &          2.30 &          2.30 &           0.06 &          0.19 &          0.58 &  \textbf{0.99} &          2.24 &           1.17 & \textbf{0.05} \\ \bottomrule
\end{tabular}
\end{table}

Moving into the results of the backdoor attack, which are detailed in Table~\ref{tab:one-client-backdoor-losstrue} (loss-based evaluation) and Table~\ref{tab:one-client-backdoor-lossfalse} (accuracy-based evaluation). Each table provides two sets of metrics: one for the primary, benign task, where high performance is the goal, and another for the secondary, adversarial task, where the desired outcome is low performance.

\rewrite{Again, we will focus on \textbf{RQ2}. A general overview of the results shows that the metrics for the original task are again similar to those in the \rewrite{no-attack} scenario. This is expected, as a key objective of a backdoor attack is to interfere as little as possible with the main learning task. When we examine the metrics for the secondary task, we see that while accuracy is reasonably high, the loss is orders of magnitude greater than the loss for the primary task. Given that the objective of a backdoor attack is to minimize loss in the secondary task, we can conclude that the attack was not successfully executed and provide the same answer for \textbf{RQ2} \rewrite{as in the labelflip attack}. This indicates that our framework also defends against this type of attack. The high accuracy, coupled with the high loss, suggests that the predictions are correct but have low confidence. This issue could easily be mitigated in a real-world application by implementing a confidence threshold, a common practice in production systems.}

However, these conclusions do not apply to FedAvg. This method shows both high accuracy and low loss in the backdoor task, making it the only schema that failed to mitigate this attack. We hypothesize that PoFL's ability to defend against this attack is due to competing gradients. By attempting to minimize both the primary and secondary tasks, the optimization for the primary task becomes less aggressive. As a result, the compromised model pool is frequently discarded because it is not considered optimal according to the evaluation metric. The \rewrite{comparison between the robustness of PoFL and FedAvg} in this scenario is presented in Figure~\ref{fig:one_backdoor_fashion_loss}.

\rewrite{On the other hand, considering \textbf{RQ4}, a closer look \rewrite{at} the primary task reveals that the overfitting issue observed with Krum and GeoMed in the labelflip attack is even more pronounced here. Consistent with our previous findings, their RFC variants, K-RFC and G-RFC, effectively reduce this overfitting.}

\begin{table}[!ht]
\centering
\tiny
\caption{One-client scenario with Backdoor attack, using loss-based evaluation.}
\label{tab:one-client-backdoor-losstrue}
\begin{tabular}{l|l|rrr|rrr|rrr|rrr}
	\toprule
	                               &            &             \multicolumn{3}{c|}{Accuracy}              &           \multicolumn{3}{c|}{Loss}           &     \multicolumn{3}{c|}{Backdoor Accuracy}      &        \multicolumn{3}{c}{Backdoor Loss}        \\
	                               &            &                  Final &          Mean &           Max &         Final &          Mean &           Min &         Final &           Mean &            Max &          Final &           Mean &           Min \\
	Dataset                        & Aggregator &                        &       Last 10 &               &               &       Last 10 &               &               &        Last 10 &                &                &        Last 10 &               \\ \midrule
	\multirow{8}{*}{Celeba-S}      & \paco{B-RFC}        &                   0.84 &          0.87 &          0.88 &          0.55 &          0.44 &          0.32 &          0.75 &           0.76 &           0.78 &           1.52 &           1.37 &          0.71 \\
	                               & Bulyan     &                   0.78 &          0.84 &          0.89 &          1.07 &          0.61 &          0.39 & \textbf{0.76} &  \textbf{0.75} &  \textbf{0.78} &           1.40 &  \textbf{1.42} &          0.79 \\
	                               & \paco{G-RFC}        &                   0.86 &          0.88 &          0.90 &          0.36 &          0.39 &          0.32 & \textbf{0.73} &           0.76 &           0.79 &           0.86 &           1.16 &          0.80 \\
	                               & GeoMed     &                   0.85 &          0.87 &          0.90 &          0.56 &          0.46 &          0.33 &          0.79 &           0.76 &           0.79 &           1.12 &           1.30 &          0.84 \\
	                               & \paco{K-RFC}        &                   0.88 &          0.88 &          0.90 &          0.45 &          0.42 &          0.29 &          0.78 &           0.77 &           0.79 &           1.54 &           1.33 &          0.74 \\
	                               & Krum       &                   0.88 &          0.87 &          0.90 &          0.60 &          0.45 &          0.31 &          0.77 & \textbf{ 0.75} &  \textbf{0.78} &  \textbf{2.01} &           1.38 & \textbf{0.89} \\
	                               & PoFL       &          \textbf{0.92} & \textbf{0.92} & \textbf{0.92} & \textbf{0.24} & \textbf{0.23} & \textbf{0.21} &          0.79 &           0.80 &           0.81 &           1.34 &           1.21 &          0.64 \\ 
	                               & FedAvg     &                   0.90 &          0.87 &          0.91 &          0.33 &          0.41 &          0.25 &          0.79 &           0.76 &           0.80 &           0.49 &           0.61 &          0.46 \\\midrule
	\multirow{8}{*}{CIFAR-10}      & \paco{B-RFC}        &                   0.85 &          0.84 &          0.86 &          0.57 &          0.56 &          0.45 &          0.61 &           0.61 &           0.65 &           3.08 &           3.24 &          1.97 \\
	                               & Bulyan     &                   0.79 &          0.79 &          0.80 &          0.76 &          0.74 &          0.69 & \textbf{0.59} &  \textbf{0.58} &  \textbf{0.60} &           3.37 &           3.24 & \textbf{2.09} \\
	                               & GeoMed     &                   0.79 &          0.79 &          0.81 &          0.71 &          0.75 &          0.63 & \textbf{0.59} &           0.59 &           0.61 &           3.18 &           3.30 &          1.87 \\
	                               & \paco{K-RFC}        &                   0.79 &          0.80 &          0.81 &          0.75 &          0.70 &          0.62 & \textbf{0.59} &           0.60 &           0.63 &           2.90 &           3.08 &          2.02 \\
	                               & Krum       &                   0.80 &          0.79 &          0.81 &          0.67 &          0.72 &          0.63 &          0.60 &           0.59 &  \textbf{0.60} &           2.98 &           3.24 &          1.85 \\
	                               & PoFL       &          \textbf{0.96} & \textbf{0.96} & \textbf{0.96} & \textbf{0.20} & \textbf{0.21} & \textbf{0.15} &          0.70 &           0.72 &           0.94 &  \textbf{5.96} &  \textbf{5.62} &          0.27 \\ 
	                               & FedAvg     &                   0.84 &          0.69 &          0.87 &          0.58 &          0.64 &          0.47 &          0.86 &           0.75 &           0.91 &           0.43 &           0.61 &          0.30 \\\midrule
	\multirow{8}{*}{Fashion MNIST} & \paco{B-RFC}        &                   0.89 &          0.89 &          0.89 &          0.40 &          0.42 &          0.36 &          0.75 &           0.74 &           0.76 &           2.53 &           2.69 &          1.67 \\
	                               & Bulyan     &                   0.88 &          0.88 &          0.88 &          0.47 &          0.47 &          0.42 &          0.73 &           0.73 &           0.74 &           3.20 &           3.09 &          1.69 \\
	                               & \paco{G-RFC}        &                   0.87 &          0.86 &          0.87 &          0.86 &          0.86 &          0.49 &          0.71 &           0.72 &           0.73 &           5.95 &           5.08 & \textbf{1.74} \\
	                               & GeoMed     &                   0.83 &          0.83 &          0.84 &          2.68 &          2.93 &          0.56 &          0.70 &           0.69 &  \textbf{0.71} &           9.57 &          12.68 &          1.53 \\
	                               & \paco{K-RFC}        &                   0.85 &          0.86 &          0.87 &          0.58 &          0.87 &          0.50 &          0.70 &           0.72 &           0.74 &           2.78 &           4.92 &          1.64 \\
	                               & Krum       &                   0.82 &          0.83 &          0.84 &          3.51 &          3.10 &          0.56 & \textbf{0.69} &  \textbf{0.69} &           0.71 & \textbf{12.68} & \textbf{13.37} &          1.72 \\
	                               & PoFL       &          \textbf{0.91} & \textbf{0.91} & \textbf{0.91} & \textbf{0.33} & \textbf{0.32} & \textbf{0.31} &          0.76 &           0.77 &           0.78 &           3.04 &           2.65 &          1.48 \\ 
	                               & FedAvg     &                   0.88 &          0.88 &          0.89 &          0.41 &          0.50 &          0.36 &          0.91 &           0.91 &           0.93 &           0.29 &           0.35 &          0.24 \\\midrule
	\multirow{8}{*}{EMNIST}        & \paco{B-RFC}        &                   0.98 &          0.98 &          0.98 &          0.13 &          0.10 &          0.07 &          0.79 &           0.80 &           0.83 &           6.59 &           5.20 &          1.82 \\
	                               & Bulyan     &                   0.97 &          0.97 &          0.98 &          0.15 &          0.13 &          0.10 &          0.78 &           0.79 &           0.80 &           6.07 &           5.96 &          1.97 \\
	                               & \paco{G-RFC}        &                   0.97 &          0.97 &          0.97 &          0.60 &          0.55 &          0.14 &          0.77 &           0.78 &           0.80 & \textbf{20.16} & \textbf{19.72} &          1.71 \\
	                               & GeoMed     &                   0.95 &          0.94 &          0.96 &          0.83 &          0.97 &          0.31 &          0.77 &           0.76 &           0.78 &          13.54 &          14.47 &          1.93 \\
	                               & \paco{K-RFC}        &                   0.97 &          0.97 &          0.97 &          0.46 &          0.46 &          0.15 &          0.78 &           0.78 &           0.80 &          16.95 &          16.52 &          1.38 \\
	                               & Krum       &                   0.95 &          0.95 &          0.95 &          0.84 &          0.94 &          0.22 & \textbf{0.76} &  \textbf{0.75} & \textbf{ 0.77} &          16.42 &          15.41 & \textbf{1.99} \\
	                               & PoFL       & \textbf{\textbf{0.99}} & \textbf{0.99} & \textbf{0.99} & \textbf{0.05} & \textbf{0.05} & \textbf{0.04} &          0.79 &           0.80 &           0.82 &           5.42 &           5.26 &          1.47 \\
	                               & FedAvg     &                   0.96 &          0.97 &          0.98 &          0.31 &          0.14 &          0.07 &          0.96 &           0.97 &           0.98 &           0.37 &           0.17 &          0.09 \\ \bottomrule
\end{tabular}
\end{table}

\begin{table}[!ht]
\centering
\tiny
\caption{One-client scenario with Backdoor attack, using accuracy-based evaluation.}
\label{tab:one-client-backdoor-lossfalse}
\begin{tabular}{l|l|rrr|rrr|rrr|rrr}
	\toprule
	                               &            &         \multicolumn{3}{c|}{Accuracy}          &            \multicolumn{3}{c|}{Loss}            &     \multicolumn{3}{c|}{Backdoor Accuracy}      &        \multicolumn{3}{c}{Backdoor Loss}        \\
	                               &            &          Final &          Mean &           Max &          Final &          Mean &            Min &         Final &           Mean &            Max &          Final &           Mean &           Min \\
	Dataset                        & Aggregator &                &       Last 10 &               &                &       Last 10 &                &               &        Last 10 &                &                &        Last 10 &               \\ \midrule
	\multirow{7}{*}{Celeba-S}      & \paco{B-RFC}        &           0.84 &          0.87 &          0.89 & \textbf{ 2.20} &          0.83 &           0.35 & \textbf{0.72} &           0.76 &           0.78 &  \textbf{3.62} &  \textbf{2.11} &          0.67 \\
	                               & Bulyan     &           0.83 &          0.80 &          0.88 &           0.46 &          0.74 &           0.38 & \textbf{0.72} &  \textbf{0.71} &  \textbf{0.77} &           0.94 &           1.36 &          0.69 \\
	                               & GeoMed     &           0.88 &          0.87 &          0.90 &           0.48 &          0.45 &           0.33 &          0.78 &           0.77 &           0.79 &           1.10 &           1.22 &          0.69 \\
	                               & \paco{K-RFC}        &           0.89 &          0.89 &          0.90 &           0.44 &          0.46 &           0.34 &          0.79 &           0.78 &           0.79 &           1.44 &           1.47 & \textbf{0.82} \\
	                               & Krum       &           0.86 &          0.86 &          0.89 &           0.61 &          0.53 &           0.33 &          0.77 &           0.76 &           0.79 &           1.16 &           1.17 &          0.75 \\
	                               & PoFL       &  \textbf{0.92} & \textbf{0.92} & \textbf{0.92} &  \textbf{0.25} & \textbf{0.30} &  \textbf{0.21} &          0.80 &           0.80 &           0.80 &           1.37 &           1.51 &          0.60 \\ 
	                               & FedAvg     &           0.87 &          0.88 & \textbf{0.92} &           0.38 &          0.41 &           0.26 &          0.77 &           0.78 &           0.80 &           0.51 &           0.59 &          0.48 \\\midrule
	\multirow{6}{*}{CIFAR-10}      & \paco{B-RFC}        &           0.84 &          0.83 &          0.85 &           0.56 &          0.56 &           0.50 &          0.61 &           0.62 &           0.64 &           3.29 &           3.34 & \textbf{2.36} \\
	                               & GeoMed     &           0.80 &          0.79 &          0.81 &           0.68 &          0.77 &           0.63 & \textbf{0.59} &           0.58 &  \textbf{0.60} &           3.09 &           3.29 &          2.17 \\
	                               & \paco{K-RFC}        &           0.79 &          0.79 &          0.81 &           0.68 &          0.74 &           0.66 &          0.60 &           0.59 &           0.61 &           3.16 &           3.43 &          2.11 \\
	                               & Krum       &           0.79 &          0.79 &          0.81 &           0.80 &          0.74 &           0.65 & \textbf{0.59} &           0.59 &           0.61 &           3.63 &           3.48 &          1.97 \\
	                               & PoFL       &  \textbf{0.96} & \textbf{0.96} & \textbf{0.96} & \textbf{ 0.20} & \textbf{0.19} & \textbf{ 0.15} &          0.71 &           0.71 &           0.73 &  \textbf{6.65} & \textbf{ 6.66} &          1.58 \\ 
	                               & FedAvg     &           0.87 &          0.40 &          0.87 &           0.53 &          0.65 &           0.48 &          0.88 &  \textbf{0.57} &           0.89 &           0.45 &           0.49 &          0.38 \\\midrule
	\multirow{7}{*}{Fashion MNIST} & \paco{B-RFC}        &           0.89 &          0.89 &          0.89 &           0.53 &          0.50 &           0.38 &          0.71 &           0.74 &           0.76 &           4.49 &           3.57 & \textbf{1.97} \\
	                               & \paco{G-RFC}        &           0.85 &          0.85 &          0.86 &           2.64 &          2.04 &           0.47 &          0.72 &           0.71 &           0.74 &          12.83 &          10.47 &          1.71 \\
	                               & GeoMed     &           0.84 &          0.84 &          0.84 &           3.14 &          2.85 &           0.61 &          0.70 &  \textbf{0.69} &  \textbf{0.71} &          14.73 &          12.68 &          1.58 \\
	                               & \paco{K-RFC}        &           0.86 &          0.86 &          0.86 &           2.34 &          2.38 &           0.50 &          0.71 &           0.72 &           0.73 &          12.51 &          11.96 &          1.80 \\
	                               & Krum       &           0.84 &          0.83 &          0.84 &           3.54 &          3.50 &           0.60 & \textbf{0.69} & \textbf{ 0.69} &           0.72 & \textbf{16.88} & \textbf{15.75} &          1.84 \\
	                               & PoFL       &  \textbf{0.91} & \textbf{0.91} & \textbf{0.91} &  \textbf{0.35} & \textbf{0.35} &  \textbf{0.33} &          0.76 &           0.76 &           0.77 &           2.82 &           2.97 &          1.81 \\ 
	                               & FedAvg     &           0.82 &          0.86 &          0.89 &           0.57 &          0.64 &           0.40 &          0.86 &           0.90 &           0.92 &           0.44 &           0.45 &          0.28 \\\midrule
	\multirow{6}{*}{EMNIST}        & \paco{B-RFC}        &           0.98 &          0.98 &          0.98 &           0.10 &          0.09 &           0.06 &          0.79 &           0.79 &           0.80 &           5.45 &           5.99 &          2.45 \\
	                               & GeoMed     &           0.94 &          0.95 &          0.95 &           1.24 &          1.01 &           0.24 & \textbf{0.74} &  \textbf{0.76} &  \textbf{0.77} & \textbf{16.24} & \textbf{14.83} &          1.83 \\
	                               & \paco{K-RFC}        &           0.96 &          0.96 &          0.97 &           0.48 &          0.56 &           0.17 &          0.78 &           0.78 &           0.80 &          13.42 &          13.97 & \textbf{2.22} \\
	                               & Krum       &           0.94 &          0.95 &          0.95 &           0.94 &          1.02 &           0.40 &          0.78 &  \textbf{0.76} &           0.78 &          13.77 &          14.72 &          1.98 \\
	                               & PoFL       & \textbf{ 0.99} & \textbf{0.99} & \textbf{0.99} &  \textbf{0.05} & \textbf{0.05} &  \textbf{0.05} &          0.81 &           0.81 &           0.82 &           5.39 &           5.39 &          1.42 \\ 
	                               & FedAvg     &           0.98 &          0.94 &          0.98 &           0.11 &          0.38 &           0.09 &          0.98 &           0.94 &           0.98 &           0.12 &           0.36 &          0.08 \\\bottomrule
\end{tabular}
\end{table}

\subsection{Multi-Client Attack Scenario}\label{sec:mult_clients}
\rewrite{Finally, we assume the strongest kind of threat model, where an attacker manages to coordinate an attack between every pool in our system. In this threat model, we analyze again a byzantine attack and a backdoor attack. This scenario will help us to fully answer \textbf{RQ3}.}

\begin{figure}
    \centering
    \begin{subfigure}[h]{0.45\textwidth}
        \centering
        \includegraphics[width=\textwidth]{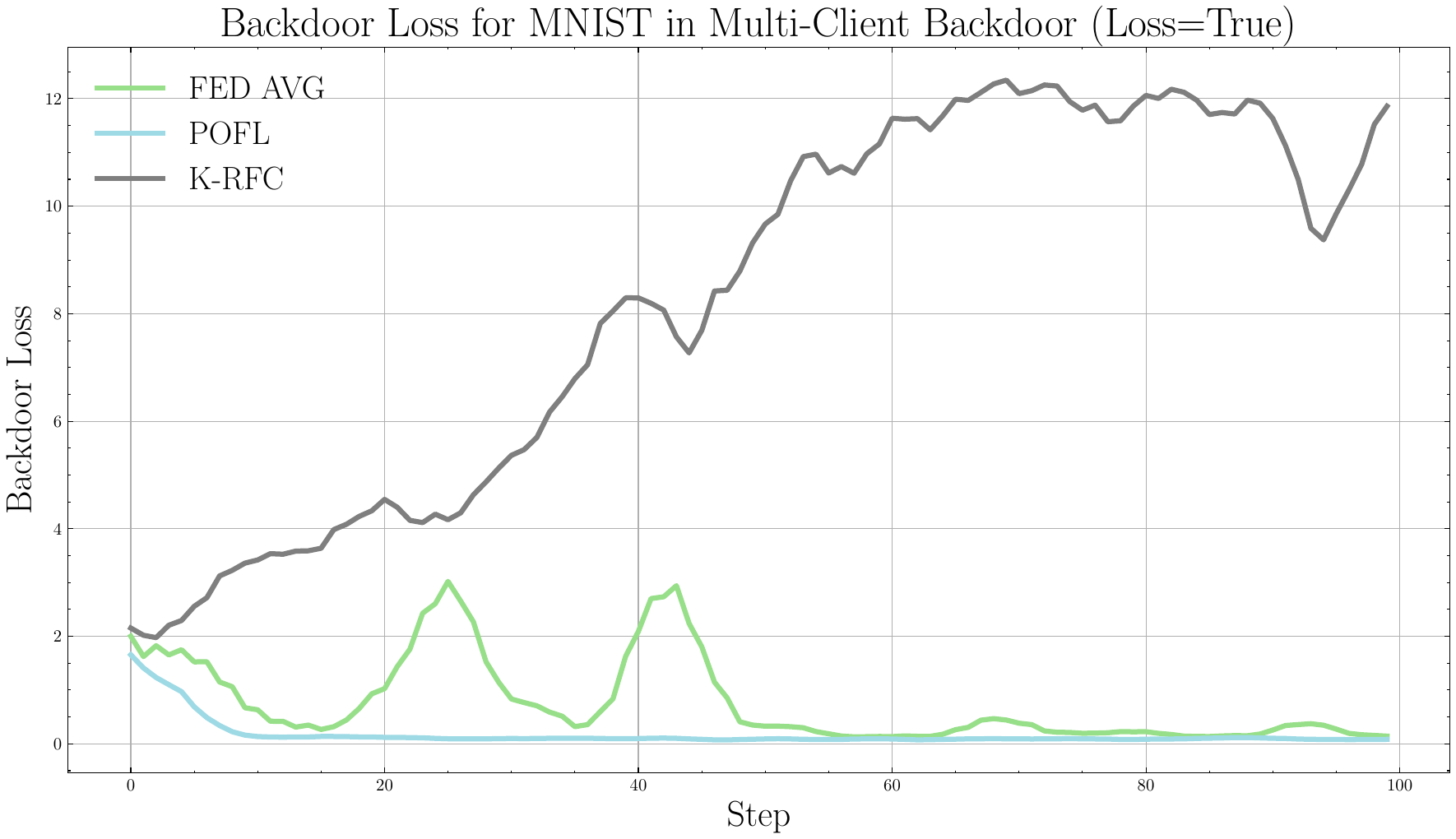}
        \caption{Backdoor loss over the rounds on MNIST dataset. Using accuracy as evaluation metric.}
        \label{fig:multi_backdoor_mnist_loss}
    \end{subfigure}
    \hfill
    \begin{subfigure}[h]{0.45\textwidth}
        \centering
        \includegraphics[width=\textwidth]{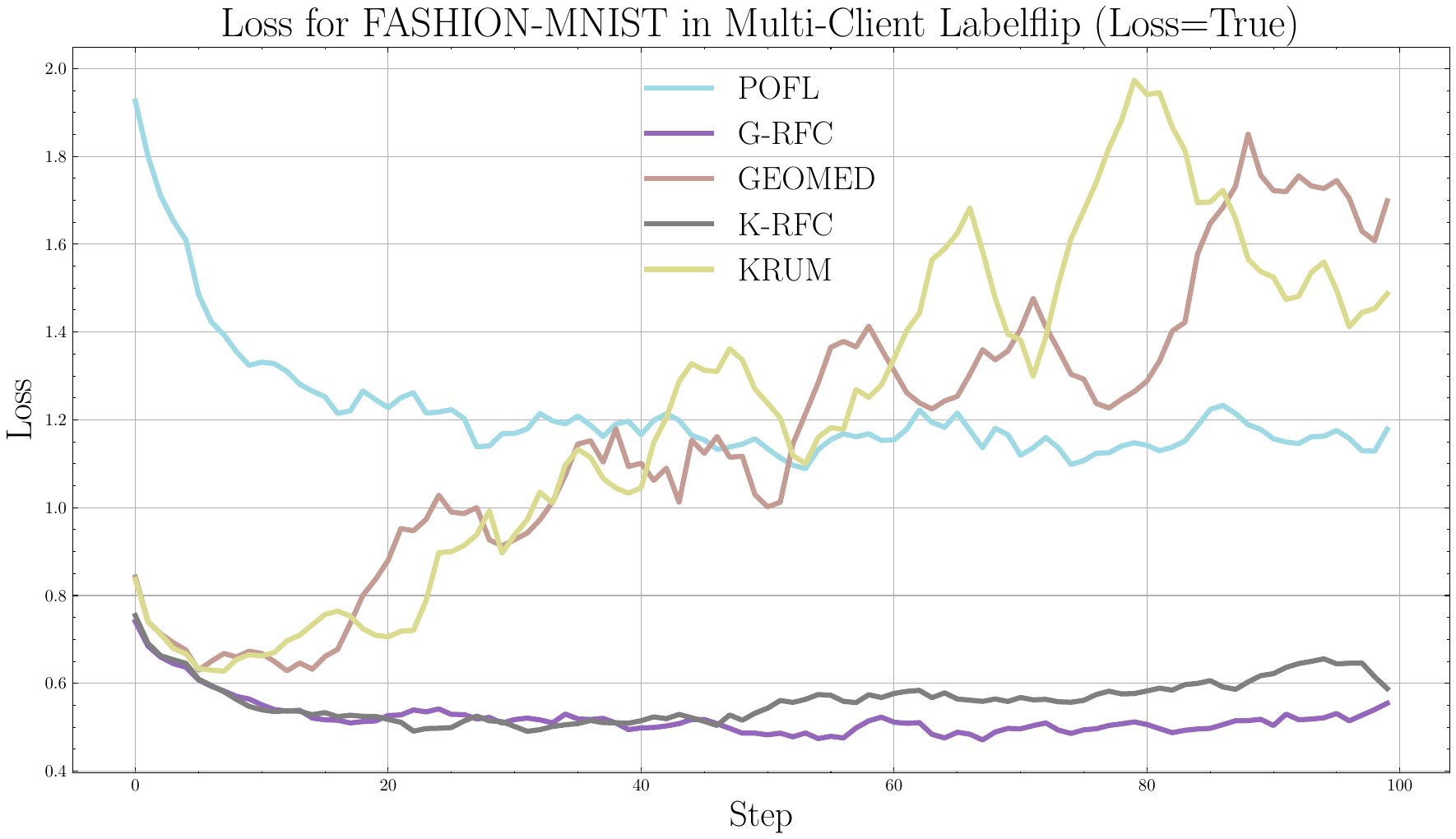}
        \caption{Labelflip loss over the rounds on Fashion-MNIST dataset. Using loss as evaluation metric.}
        \label{fig:multi_labelflip_fashionmnist_loss}
    \end{subfigure}

\caption{Performance of the proposal under the multiple attack scenario. In Figure \ref{fig:multi_backdoor_mnist_loss}, we observe how \paco{K-RFC} is able to mitigate the backdoor attack, while PoFL and FedAvg fails to do so. In Figure \ref{fig:multi_labelflip_fashionmnist_loss}, we see how \paco{RFC} is able to keep the loss low despite the presence of labelflip attacks, while other robust aggregators fail to do so.}
\label{fig:multi_attack}
\end{figure}

\rewrite{We present the results for the Byzantine attack in Table~\ref{tab:multi-client-labelflip}. For \textbf{RQ3}, our analysis shows that the PoFL framework fails to mitigate this attack effectively, which we attribute to the absence of a robust aggregation mechanism within each pool, \rewrite{highlighting} the need for \rewrite{exchangeable aggregators}. This deficiency leads to a significant degradation in performance compared to previous scenarios. The severe degradation seen in FedAvg is now also present in PoFL for the CelebA-S dataset, with both methods achieving only around 50\% accuracy. In contrast, the RFC variants do not exhibit this problem, which confirms that this new framework represents a significant improvement over the original PoFL architecture. The performance of the RFC variants remains consistent with our earlier findings, demonstrating their resilience and robustness against this severe threat model, even mitigating the overfitting seen in the \rewrite{non-RFC} variants. We present the evolution of the loss in the Fashion-MNIST dataset under \rewrite{these} conditions in Figure~\ref{fig:multi_labelflip_fashionmnist_loss} as an illustrative example of the observed \rewrite{behavior}.}

\begin{table}[!ht]
\centering
\tiny
\caption{Multi-client scenario with Labelflip attack: a comparison of loss-based and accuracy-based evaluations.}
\label{tab:multi-client-labelflip}
\begin{tabular}{l|l|rrr|rrr|rrr|rrr}
	\toprule
	           \multicolumn{2}{c|}{}            &                                   \multicolumn{6}{c|}{Loss based}                                   &                                 \multicolumn{6}{c}{Accuracy based}                                 \\ \cmidrule
	(lr){3-8} \cmidrule(lr){9-14}  &            &          \multicolumn{3}{c|}{Accuracy}           &            \multicolumn{3}{c|}{Loss}             &          \multicolumn{3}{c|}{Accuracy}          &             \multicolumn{3}{c}{Loss}             \\
	                               &            &         Final &           Mean &             Max &          Final &           Mean &            Min &          Final &          Mean &            Max &          Final &           Mean &            Min \\ \midrule
	Dataset                        & Aggregator &               &        Last 10 &                 &                &        Last 10 &                &                &       Last 10 &                &                &        Last 10 &                \\
	\multirow{8}{*}{Celeba-S}      & B-RFC        &          0.84 &           0.84 &            0.88 &           0.60 &           0.55 &           0.39 &           0.87 &           0.88 &  \textbf{0.90} &           0.48 &           0.64 &           0.40 \\
	                               & Bulyan     &          0.73 &           0.83 &            0.89 &           1.36 &           0.70 &           0.42 &  \textbf{0.89} &           0.84 &           0.89 &           0.37 &           0.53 &           0.37 \\
	                               & G-RFC        &          0.88 &           0.87 &            0.89 & \textbf{ 0.34} &           0.40 &  \textbf{0.31} &  \textbf{0.89} &           0.88 &  \textbf{0.90} & \textbf{ 0.41} &  \textbf{0.46} &  \textbf{0.32} \\
	                               & GeoMed     &          0.80 &           0.86 &  \textbf{ 0.90} &           0.72 &           0.51 &           0.34 &  \textbf{0.89} &           0.86 &           0.89 &  \textbf{0.35} &           0.50 &           0.35 \\
	                               & K-RFC        & \textbf{0.89} &  \textbf{0.88} &            0.89 &           0.46 & \textbf{ 0.38} &  \textbf{0.31} &  \textbf{0.89} &  \textbf{0.89} &           0.89 &           0.50 &           0.47 &           0.33 \\
	                               & Krum       &          0.84 &           0.86 &            0.89 &           0.63 &           0.45 &           0.33 &           0.88 &           0.83 &           0.89 &           0.42 &           0.51 &           0.33 \\
	                               & PoFL       &          0.51 &           0.49 &            0.53 &           1.27 &           1.26 &           0.87 &           0.52 &           0.51 &           0.58 &           1.35 &           1.97 &           0.75 \\ 
	                               & FedAvg     &          0.29 &           0.39 &            0.55 &           1.47 &           1.34 &           0.95 &           0.31 &           0.36 &           0.66 &           1.44 &           1.49 &           0.65 \\\midrule
	\multirow{8}{*}{CIFAR-10}      & \paco{B-RFC}        &          0.80 &           0.80 &            0.80 &           0.70 &           0.71 &           0.65 &           0.80 &          0.80 &           0.80 &           0.75 &           0.71 &           0.64 \\
	                               & Bulyan     &          0.83 &           0.83 &            0.84 &           0.60 &           0.59 &           0.54 &           0.84 &          0.83 &           0.84 &           0.56 &           0.58 &           0.56 \\
	                               & \paco{G-RFC}        & \textbf{0.84} &  \textbf{0.84} &   \textbf{0.85} & \textbf{ 0.54} &  \textbf{0.54} &  \textbf{0.52} &  \textbf{0.85} & \textbf{0.84} &  \textbf{0.85} &  \textbf{0.54} &  \textbf{0.56} &  \textbf{0.53} \\
	                               & GeoMed     &          0.83 &           0.83 &            0.84 &           0.58 &           0.56 & \textbf{ 0.52} &           0.84 & \textbf{0.84} &           0.84 &  \textbf{0.54} &  \textbf{0.56} &           0.54 \\
	                               & \paco{K-RFC}        & \textbf{0.84} & \textbf{ 0.84} &            0.84 &           0.55 &           0.56 &           0.53 &           0.84 & \textbf{0.84} &           0.84 &           0.58 &           0.57 &  \textbf{0.53} \\
	                               & Krum       & \textbf{0.84} & \textbf{ 0.84} &            0.84 &           0.56 &           0.55 &  \textbf{0.52} &           0.84 & \textbf{0.84} &           0.84 & \textbf{ 0.54} &  \textbf{0.56} &  \textbf{0.53} \\
	                               & PoFL       &          0.16 &           0.20 &            0.40 &           2.17 &           2.10 &           1.76 &           0.11 &          0.17 &           0.35 &           3.96 &           2.96 &           2.15 \\ 
	                               & FedAvg     &          0.10 &           0.12 &            0.36 &            NaN &           2.39 &           2.15 &           0.10 &          0.12 &           0.34 &            NaN &           2.48 &           2.12 \\\midrule
	\multirow{8}{*}{Fashion MNIST} & \paco{B-RFC}        & \textbf{0.88} &  \textbf{0.88} &            0.88 &  \textbf{0.45} &  \textbf{0.44} &           0.41 &  \textbf{0.88} & \textbf{0.88} &           0.88 &  \textbf{0.44} &           0.46 &           0.43 \\
	                               & Bulyan     & \textbf{0.88} &  \textbf{0.88} &   \textbf{0.89} &           0.47 &           0.47 &           0.39 &  \textbf{0.88} & \textbf{0.88} &  \textbf{0.89} &           0.46 &  \textbf{0.45} &           0.40 \\
	                               & \paco{G-RFC}        & \textbf{0.88} &  \textbf{0.88} &  \textbf{ 0.89} &           0.61 &           0.54 &           0.43 &           0.87 &          0.87 &           0.87 &           1.26 &           1.20 &           0.50 \\
	                               & GeoMed     &          0.85 &           0.85 &            0.86 &           1.96 &           1.71 &           0.59 &           0.85 &          0.85 &           0.85 &           2.36 &           1.89 &           0.58 \\
	                               & \paco{K-RFC}        & \textbf{0.88} &  \textbf{0.88} &            0.88 &           0.57 &           0.62 &           0.46 &           0.86 &          0.86 &           0.87 &           1.49 &           1.53 &           0.53 \\
	                               & Krum       &          0.84 &           0.85 &            0.85 &           1.69 &           1.52 &           0.58 &           0.86 &          0.85 &           0.86 &           1.91 &           1.65 &           0.57 \\
	                               & PoFL       &          0.79 &           0.82 &            0.85 &           1.38 &           1.17 &           0.97 &           0.83 &          0.83 &           0.85 &           1.21 &           1.19 &           1.03 \\ 
	                               & FedAvg     &          0.32 &           0.59 & \textbf{ 0.89 } &           2.24 &           1.30 & \textbf{ 0.32} &           0.14 &          0.56 & \textbf{ 0.89} &           2.30 &           1.31 & \textbf{ 0.31} \\\midrule
	\multirow{8}{*}{EMNIST}        & \paco{B-RFC}        &          0.97 &  \textbf{0.98} &            0.98 &  \textbf{0.11} &  \textbf{0.10} &           0.08 &           0.97 &          0.97 &  \textbf{0.98} &           0.13 &           0.12 &           0.09 \\
	                               & Bulyan     & \textbf{0.98} &  \textbf{0.98} &            0.98 &  \textbf{0.11} & \textbf{ 0.10} &           0.08 &  \textbf{0.98} & \textbf{0.98} &  \textbf{0.98} & \textbf{ 0.10} &  \textbf{0.10} &           0.08 \\
	                               & \paco{G-RFC}        &          0.97 &           0.97 &            0.98 &           0.23 &           0.23 &           0.12 &           0.97 &          0.97 &  \textbf{0.98} &           0.35 &           0.32 &           0.13 \\
	                               & GeoMed     &          0.97 &           0.96 &            0.97 &           0.59 &           0.58 &           0.19 &           0.96 &          0.96 &           0.97 &           0.69 &           0.60 &           0.19 \\
	                               & \paco{K-RFC}        &          0.97 &           0.97 &            0.98 &           0.23 &           0.21 &           0.12 &           0.97 &          0.97 &  \textbf{0.98} &           0.32 &           0.31 &           0.14 \\
	                               & Krum       &          0.96 &           0.96 &            0.97 &           0.60 &           0.59 &           0.23 &           0.96 &          0.96 &           0.97 &           0.53 &           0.53 &           0.17 \\
	                               & PoFL       &          0.94 &           0.91 &            0.96 &           1.09 &           1.12 &           0.81 &           0.95 &          0.95 &           0.96 &           0.93 &           1.04 &           0.57 \\ 
	                               & FedAvg     & \textbf{0.98} &           0.76 &   \textbf{0.99} &           0.13 &           1.11 &  \textbf{0.05} &           0.65 &          0.63 &           0.98 &           2.18 &           1.16 &  \textbf{0.06} \\\bottomrule
\end{tabular}
\end{table}

An examination of the backdoor attack, as detailed in Tables \ref{tab:multi-client-backdoor-losstrue} and \ref{tab:multi-client-backdoor-lossfalse}, reveals a pattern consistent with our observations in the Byzantine attack scenario. The PoFL framework demonstrates a clear failure to resist this attack. This is evidenced by the model's successful learning of malicious tasks on datasets like EMNIST and CIFAR-10. For the Celeba-S dataset, while the primary task performance is maintained, PoFL still fails to mitigate the backdoor, as reflected by the low backdoor loss values. We attribute this failure to the model replacement technique inherent to the PoFL framework, which renders it unsuitable for mitigating this type of threat.

In contrast, our proposed RFC variants successfully mitigate the backdoor attack. They achieve this in a manner similar to the single pool compromise scenario, effectively neutralizing the attack while simultaneously reducing the overfitting effect often associated with certain robust aggregation techniques. These results validate our proposal as a robust and effective defense mechanism against backdoor attacks. We present an illustration of this phenomena in Figure~\ref{fig:multi_backdoor_mnist_loss}.

\rewrite{Thus, we can conclude that the answer to \textbf{RQ3} is \rewrite{yes}: the hierarchical combination of \rewrite{intra-pool} robust aggregation and \rewrite{blockchain-level} consensus can withstand a coordinated \rewrite{multi-pool} attack where the majority of clusters are compromised, rendering RFC a powerful defense mechanism while avoiding multiple drawbacks common to centralized FL, such as \rewrite{lack of incentivization} or single points of failure.}
\begin{table}[!ht]
\centering
\tiny
\caption{Multi-client scenario with Backdoor attack, using loss-based evaluation.}
\label{tab:multi-client-backdoor-losstrue}
\begin{tabular}{l|l|rrr|rrr|rrr|rrr}
	\toprule
	                               &            &          \multicolumn{3}{c|}{Accuracy}          &            \multicolumn{3}{c|}{Loss}            &     \multicolumn{3}{c|}{Backdoor Accuracy}     &        \multicolumn{3}{c}{Backdoor Loss}         \\
	                               &            &          Final &          Mean &            Max &          Final &           Mean &           Min &         Final &           Mean &           Max &          Final &            Mean &           Min \\
	Dataset                        & Aggregator &                &       Last 10 &                &                &        Last 10 &               &               &        Last 10 &               &                &         Last 10 &               \\ \midrule
	\multirow{8}{*}{Celeba-S}      & \paco{B-RFC}        &           0.85 & \textbf{0.86} &  \textbf{0.89} &           0.42 &           0.72 &          0.36 &          0.77 &           0.76 &          0.79 &           1.23 &            1.41 &          0.74 \\
	                               & Bulyan     &           0.85 &          0.83 &           0.87 &           0.42 &           0.65 &          0.42 &          0.72 &  \textbf{0.73} & \textbf{0.78} &           1.20 &  \textbf{1.55} & \textbf{0.76} \\
	                               & \paco{G-RFC}        &           0.84 &          0.85 &           0.90 &           0.41 &           0.43 &          0.31 &          0.75 &           0.74 &          0.79 &           1.04 &            1.14 &          0.68 \\
	                               & GeoMed     &           0.86 &          0.86 &           0.89 &           0.55 &           0.53 &          0.31 &          0.77 &           0.75 &          0.79 &           1.14 &            1.33 &          0.73 \\
	                               & \paco{K-RFC}        &           0.89 &          0.87 &           0.89 &           0.36 &           0.40 &          0.32 &          0.78 &           0.76 &          0.79 & \textbf{ 1.25} &            1.14 &          0.70 \\
	                               & Krum       &           0.80 &          0.84 &           0.89 &           0.60 &           0.53 &          0.35 & \textbf{0.69} &           0.74 & \textbf{0.78} &           1.17 &            1.19 &          0.71 \\
	                               & PoFL       &           0.90 &          0.90 &           0.91 &           0.31 & \textbf{0.31} & \textbf{0.26} &          0.79 &           0.79 &          0.80 &           0.54 &            0.57 &          0.47 \\ 
	                               & FedAvg     &  \textbf{0.91} & \textbf{0.91} &  \textbf{0.92} &  \textbf{0.29} &           0.34 &          0.23 &          0.80 &           0.80 &          0.81 &           0.62 &            0.55 &          0.45 \\\midrule
	\multirow{7}{*}{CIFAR-10}      & \paco{B-RFC}        &           0.84 &          0.84 &           0.85 &           0.54 &           0.56 &          0.52 &          0.61 &           0.62 &          0.65 &           3.08 &            3.20 & \textbf{2.34} \\
	                               & Bulyan     &           0.78 &          0.78 &           0.80 &           0.77 &           0.75 &          0.69 & \textbf{0.58} &  \textbf{0.58} & \textbf{0.59} &           3.32 &            3.19 &          2.24 \\
	                               & GeoMed     &           0.80 &          0.79 &           0.81 &           0.68 &           0.71 &          0.62 &          0.59 &           0.59 &          0.61 &           3.39 &            3.28 &          2.15 \\
	                               & \paco{K-RFC}        &           0.80 &          0.80 &           0.81 &           0.72 &           0.69 &          0.64 & \textbf{0.58} &           0.59 &          0.62 &           3.56 &   \textbf{3.41} &          2.22 \\
	                               & Krum       &           0.78 &          0.79 &           0.81 &           0.75 &           0.71 &          0.66 & \textbf{0.58} &           0.59 &          0.61 &  \textbf{3.90} &            3.34 &          2.08 \\
	                               & PoFL       &           0.86 &          0.78 &           0.87 &           0.50 &           0.53 &          0.45 &          0.88 &           0.84 &          0.91 &           0.38 &            0.39 &          0.30 \\ 
	                               & FedAvg     & \textbf{ 0.93} & \textbf{0.92} &  \textbf{0.94} &  \textbf{0.28} &  \textbf{0.34} & \textbf{0.24} &          0.93 &           0.94 &          0.96 &           0.25 &            0.25 &          0.16 \\\midrule
	\multirow{8}{*}{Fashion MNIST} & \paco{B-RFC}        &  \textbf{0.89} & \textbf{0.89} &           0.89 &  \textbf{0.40} &  \textbf{0.40} &          0.36 &          0.74 &           0.74 &          0.75 &           2.61 &            2.69 &          1.74 \\
	                               & Bulyan     &           0.88 &          0.88 &           0.88 &           0.45 &           0.47 &          0.41 &          0.73 &           0.73 &          0.74 &           2.75 &            2.95 & \textbf{1.90} \\
	                               & \paco{G-RFC}        &           0.86 &          0.86 &           0.87 &           0.92 &           0.97 &          0.50 &          0.74 &           0.71 &          0.74 &           4.45 &            5.11 &          1.61 \\
	                               & GeoMed     &           0.84 &          0.83 &           0.84 &           2.77 &           3.00 &          0.57 &          0.71 &           0.69 & \textbf{0.71} & \textbf{12.71} & \textbf{ 13.84} &          1.71 \\
	                               & \paco{K-RFC}        &           0.86 &          0.86 &           0.87 &           0.75 &           0.97 &          0.43 &          0.72 &           0.71 &          0.74 &           3.99 &            5.19 &          1.62 \\
	                               & Krum       &           0.81 &          0.82 &           0.84 &           3.77 &           2.89 &          0.60 & \textbf{0.69} &  \textbf{0.68} & \textbf{0.71} &          12.39 &           12.10 &          0.90 \\
	                               & PoFL       &           0.87 &          0.88 &           0.89 &           0.61 &           0.51 &          0.38 &          0.89 &           0.91 &          0.92 &           0.48 &            0.37 &          0.28 \\ 
	                               & FedAvg     &           0.87 & \textbf{0.89} &  \textbf{0.90} &           0.41 &           0.54 & \textbf{0.31} &          0.91 &           0.92 &          0.94 &           0.30 &            0.39 &          0.21 \\\midrule
	\multirow{8}{*}{EMNIST}        & \paco{B-RFC}        &  \textbf{0.98} & \textbf{0.98} &  \textbf{0.98} &  \textbf{0.07} & \textbf{ 0.08} & \textbf{0.06} &          0.80 &           0.80 &          0.82 &           4.22 &            4.67 &          1.84 \\
	                               & Bulyan     &           0.97 &          0.97 &  \textbf{0.98} &           0.13 &           0.11 &          0.09 &          0.79 &           0.79 &          0.80 &           5.19 &            4.61 & \textbf{2.08} \\
	                               & \paco{G-RFC}        &           0.97 &          0.97 &           0.97 &           0.41 &           0.36 &          0.13 &          0.79 &           0.78 &          0.80 & \textbf{13.90} &           12.68 &          1.75 \\
	                               & GeoMed     &           0.94 &          0.94 &           0.96 &           1.16 &           1.08 &          0.24 & \textbf{0.75} & \textbf{ 0.75} & \textbf{0.77} &          13.71 &  \textbf{13.74} &          1.79 \\
	                               & \paco{K-RFC}        &           0.97 &          0.97 &           0.97 &           0.29 &           0.29 &          0.13 &          0.78 &           0.78 &          0.79 &          12.06 &           10.62 &          1.88 \\
	                               & Krum       &           0.95 &          0.95 &           0.96 &           0.74 &           0.75 &          0.24 &          0.76 &           0.77 &          0.79 &          13.91 &           13.45 &          1.86 \\
	                               & PoFL       &  \textbf{0.98} &          0.98 &  \textbf{0.98} &  \textbf{0.07} &  \textbf{0.08} & \textbf{0.06} &          0.98 &           0.98 &          0.99 &           0.09 &            0.08 &          0.06 \\ 
	                               & FedAvg     &  \textbf{0.98} &          0.96 & \textbf{ 0.98} &  \textbf{0.07} &           0.19 &          0.07 &          0.98 &           0.95 &          0.98 &           0.10 &            0.24 &          0.10 \\\bottomrule
\end{tabular}
\end{table}

\begin{table}[!ht]
\centering
\tiny
\caption{Multi-client scenario with Backdoor attack, using accuracy-based evaluation.}
\label{tab:multi-client-backdoor-lossfalse}
\begin{tabular}{l|l|rrr|rrr|rrr|rrr}
	\toprule
	                               &            &          \multicolumn{3}{c|}{Accuracy}           &            \multicolumn{3}{c|}{Loss}            &     \multicolumn{3}{c|}{Backdoor Accuracy}      &        \multicolumn{3}{c}{Backdoor Loss}        \\
	                               &            &          Final &           Mean &            Max &         Final &           Mean &            Min &         Final &           Mean &            Max &          Final &           Mean &           Min \\
	Dataset                        & Aggregator &                &        Last 10 &                &               &        Last 10 &                &               &        Last 10 &                &                &        Last 10 &               \\ \midrule
	\multirow{8}{*}{Celeba-S}      & \paco{B-RFC}        &           0.88 &           0.87 &           0.90 &           0.44 &           0.83 &           0.36 &          0.77 &           0.76 &          0.79 &           1.38 &  \textbf{1.97} &          0.86 \\
	                               & Bulyan     &           0.80 &           0.83 &           0.89 &           1.45 &           0.68 &           0.42 & \textbf{0.73} &  \textbf{0.74} & \textbf{0.78} &  \textbf{1.88} &           1.49 & \textbf{0.88} \\
	                               & \paco{G-RFC}        &           0.90 &           0.88 &           0.91 &           0.36 &           0.40 &           0.27 &          0.77 &           0.76 &          0.80 &           1.61 &           1.39 &          0.60 \\
	                               & GeoMed     &           0.86 &           0.87 &           0.89 &           0.42 &           0.42 &           0.33 & \textbf{0.73} &           0.76 &          0.80 &           1.21 &           1.27 &          0.78 \\
	                               & \paco{K-RFC}        &           0.90 &           0.89 &           0.90 &           0.46 &           0.50 &           0.34 &          0.79 &           0.77 &          0.79 &           1.66 &           1.68 &          0.72 \\
	                               & Krum       &           0.87 &           0.86 &           0.90 &           0.43 &           0.50 &           0.33 &          0.74 &           0.76 &          0.79 &           1.53 &           1.32 &          0.78 \\
	                               & FedAvg     &  \textbf{0.92} &  \textbf{0.91} &  \textbf{0.92} &  \textbf{0.29} &  \textbf{0.30} &  \textbf{0.23} &          0.80 &           0.79 &          0.81 &           0.49 &           0.54 &          0.46 \\ 
	                               & PoFL       &           0.89 &           0.90 &           0.91 &           0.51 &           0.37 &           0.28 &          0.79 &           0.79 &          0.80 &           0.64 &           0.56 &          0.46 \\ \midrule
	\multirow{6}{*}{CIFAR-10}      & \paco{B-RFC}        &           0.84 &           0.84 &           0.86 &          0.53 &           0.56 &           0.49 &          0.63 &           0.62 &           0.64 &  \textbf{3.53} &           3.53 &          2.05 \\
	                               & GeoMed     &           0.81 &           0.80 &           0.82 &          0.67 &           0.67 &           0.63 &          0.58 &           0.60 &  \textbf{0.62} & \textbf{ 3.46} &           3.03 &          1.87 \\
	                               & \paco{K-RFC}        &           0.80 &           0.80 &           0.82 &          0.66 &           0.70 &           0.63 &          0.60 &           0.60 &  \textbf{0.62} &           3.26 &  \textbf{3.53} & \textbf{2.18} \\
	                               & Krum       &           0.79 &           0.80 &           0.81 &          0.71 &           0.68 &           0.63 & \textbf{0.56} &  \textbf{0.59} &  \textbf{0.62} &           3.24 &           3.25 &          1.73 \\
	                               & PoFL       &           0.82 &           0.75 &           0.87 &          0.74 &           0.63 &           0.45 &          0.87 &           0.82 &           0.91 &           0.49 &           0.46 &          0.33 \\ 
	                               & FedAvg     &  \textbf{0.93} & \textbf{ 0.92} &  \textbf{0.94} & \textbf{0.32} &  \textbf{0.34} &  \textbf{0.27} &          0.94 &           0.94 &           0.95 &           0.22 &           0.25 &          0.19 \\ \midrule
	\multirow{6}{*}{Fashion MNIST} & \paco{B-RFC}        &  \textbf{0.88} &  \textbf{0.89} &           0.89 &          0.48 &  \textbf{0.47} &           0.38 &          0.75 &           0.74 &           0.76 &           3.18 &           3.38 &          1.60 \\
	                               & GeoMed     &           0.84 &           0.84 &           0.85 &          3.99 &           3.50 &           0.54 & \textbf{0.68} & \textbf{ 0.70} &           0.72 & \textbf{15.44} & \textbf{13.74} & \textbf{1.81} \\
	                               & \paco{K-RFC}        &           0.86 &           0.86 &           0.86 &          2.03 &           2.01 &           0.53 &          0.73 &           0.72 &           0.74 &          11.06 &          11.08 &          1.74 \\
	                               & Krum       &           0.83 &           0.84 &           0.84 &          2.95 &           2.80 &           0.55 &          0.69 &  \textbf{0.70} &  \textbf{0.71} &          14.83 &          13.41 &          1.56 \\
	                               & PoFL       &  \textbf{0.88} &           0.88 &           0.89 &          0.54 &           0.53 &           0.37 &          0.91 &           0.92 &           0.92 &           0.38 &           0.36 &          0.26 \\ 
	                               & FedAvg     &  \textbf{0.88} &  \textbf{0.89} &  \textbf{0.90} & \textbf{0.37} &           0.53 &  \textbf{0.31} &          0.91 &           0.92 &           0.93 &           0.26 &           0.40 &          0.22 \\\midrule
	\multirow{6}{*}{EMNIST}        & \paco{B-RFC}        &  \textbf{0.98} &  \textbf{0.98} &  \textbf{0.98} &          0.12 &  \textbf{0.09} & \textbf{ 0.06} &          0.79 &           0.80 &           0.82 &           5.82 &           5.86 & \textbf{2.19} \\
	                               & GeoMed     &           0.95 &           0.94 &           0.95 &          0.87 &           0.96 &           0.24 &          0.74 &  \textbf{0.74} & \textbf{ 0.77} &          14.94 &          14.00 &          2.03 \\
	                               & \paco{K-RFC}        &           0.97 &           0.97 &           0.97 &          0.53 &           0.52 &           0.15 &          0.78 &           0.78 &           0.79 &          15.94 & \textbf{16.35} &          1.88 \\
	                               & Krum       &           0.93 &           0.94 &           0.95 &          1.32 &           1.05 &           0.28 & \textbf{0.72} &           0.75 &  \textbf{0.77} & \textbf{18.58} &          15.35 &          1.86 \\
	                               & PoFL       & \textbf{ 0.98} & \textbf{ 0.98} &  \textbf{0.98} &          0.14 &           0.14 &           0.08 &          0.98 &           0.98 &           0.98 &           0.14 &           0.16 &          0.08 \\ 
	                               & FedAvg     &           0.97 &           0.96 &  \textbf{0.98} & \textbf{0.11} &           0.19 &  \textbf{0.06} &          0.97 &           0.96 &           0.98 &           0.10 &           0.20 &          0.08 \\
                                   \bottomrule
\end{tabular}
\end{table}

\newpage

\section{Discussion and Conclusions}\label{sec:conclusions}

\rewrite{In this work, we have presented RFC, a novel \rewrite{active defense layer} framework that bridges the gap between decentralized consensus and robust FL. By leveraging the foundational principles of PoFL and Blockchain technology, RFC addresses the critical vulnerabilities of centralized FL, namely the single point of failure and the lack of verifiable audit trails. The resulting architecture not only facilitates enhanced scalability but also establishes a secure, immutable ledger for model updates, providing the transparency essential for modern TAI systems. }

\rewrite{A central finding of our research is that the operational constraints of RFC (specifically the strict isolation of client pools) are not merely incidental limitations but are, in fact, deliberate design choices that prioritize system integrity. While this independent pool structure ensures that an adversarial update is localized and discarded by the consensus mechanism before it can poison the global state, it does introduce a trade-off regarding the effective sample size per round. In environments characterized by extreme non-IID data distributions, the exclusion of a compromised pool may temporarily reduce the statistical diversity of the training round. However, we argue that this reduction in data throughput is a necessary and advantageous compromise for maintaining \rewrite{high-fidelity} global models in hostile environments. Furthermore, the inherent flexibility of the evaluation metric $\mathcal{E}$ ensures that the framework remains \rewrite{future-proof}. By treating the consensus criterion as a modular hyperparameter, RFC can be seamlessly adapted to domain-specific utility functions beyond standard accuracy, such as fairness-weighted scores or differential privacy budgets.}

\rewrite{The empirical results obtained throughout our evaluation demonstrate that RFC significantly outperforms both standard FedAvg and foundational PoFL implementations under Byzantine and backdoor attacks. Notably, the framework’s ability to mitigate overfitting, often a \rewrite{byproduct} of using robust aggregation rules in isolation, suggests that the hierarchical combination of \rewrite{pool-level} aggregation and \rewrite{blockchain-level} consensus, constituting an \rewrite{active defense layer}, provides a stabilizing effect on the global model’s convergence. Despite these promising results, the development of RFC is an ongoing process. An opportunity for future investigation lies in the formulation of adaptive consensus mechanisms. By evolving the evaluation metric into a dynamic function that adjusts its sensitivity based on the training epoch, we can develop a \rewrite{context-aware} defense mechanism capable of neutralizing sophisticated, \rewrite{otherwise} unobserved attack vectors.}

\rewrite{Looking ahead, the next frontier for RFC lies at the intersection of robustness and the broader pillars of TAI, specifically the tension between accuracy, fairness, and transparency. Future work will focus on characterizing the Pareto frontiers that delineate the optimal \rewrite{trade-offs} among these competing objectives via the evaluation metric. Such a theoretical and empirical framework will empower practitioners to calibrate the RFC architecture for their specific operational contexts, ranging from \rewrite{high-security} financial systems where integrity is paramount, to \rewrite{fairness-critical} medical diagnostics where representation and equity are the primary goals. Ultimately, RFC provides a scalable and secure foundation upon which the next generation of decentralized, collaborative machine learning systems can be built.}

\section*{Acknowledgments}
This research is part of the Project “Ethical, Responsible and General Purpose Artificial Intelligence: Applications In Risk Scenarios” (IAFER) Exp.:TSI-100927-2023-1 funded through the Creation of university-industry research programs (Enia Programs), aimed at the research and development of artificial intelligence, for its dissemination and education within the framework of the Recovery, Transformation and Resilience Plan from the European Union Next Generation EU through the Ministry for Digital Transformation and the Civil Service.
\bibliographystyle{unsrt}  
\bibliography{references}

\end{document}